\documentclass[trackchanges]{aastex701}
\usepackage{graphicx}
\usepackage{subcaption}
\usepackage{xcolor}
\usepackage{siunitx}
\usepackage{lineno}
\usepackage{multirow}
\usepackage{array}
\usepackage{amsmath}

\begin{document}

\title{Identifying Kilonovae in the Presence of Optical Afterglow for the Wide Field Survey Telescope}

\author[0009-0008-1698-3319,gname=Huan,sname=Yang]{Huan Yang}
\affiliation{Department of Astronomy, University of Science and Technology of China, Hefei 230026, China}
\affiliation{School of Astronomy and Space Sciences, University of Science and Technology of China, Hefei 230026, China}
\email[show]{yanghuanhuan@mail.ustc.edu.cn}  
 
\author[0000-0002-2242-1514,gname=Zhengyan,sname=Liu]{Zhengyan Liu}
\affiliation{Department of Astronomy, University of Science and Technology of China, Hefei 230026, China}
\affiliation{School of Astronomy and Space Sciences, University of Science and Technology of China, Hefei 230026, China}
\email[show]{ustclzy@mail.ustc.edu.cn}

\author[0000-0004-1760-2614]{Liang-Duan Liu}
\affiliation{Institute of Astrophysics, Central China Normal University, Wuhan 430079, China}
\affiliation{Laboratory for Compact Object Astrophysics and Astronomical Technology, Central China Normal University, Wuhan 430079, China}
\affiliation{Education Research and Application Center, National Astronomical Data Center, Wuhan 430079, China}
\email[]{liuld@ccnu.edu.cn}

\author[0000-0002-1330-2329]{Wen Zhao}
\affiliation{Department of Astronomy, University of Science and Technology of China, Hefei 230026, China}
\affiliation{School of Astronomy and Space Sciences, University of Science and Technology of China, Hefei 230026, China}
\affiliation{College of Physics, Guizhou University, Guiyang 550025, People’s Republic of China}
\email[show]{wzhao7@ustc.edu.cn}

\author[0000-0002-7835-8585]{Zi-Gao Dai}
\affiliation{Department of Astronomy, University of Science and Technology of China, Hefei 230026, China}
\affiliation{School of Astronomy and Space Sciences, University of Science and Technology of China, Hefei 230026, China}
\email[]{daizg@ustc.edu.cn}

\begin{abstract}
Identifying kilonovae associated with binary neutron star mergers is often complicated by the presence of a dominant synchrotron afterglow. In this work, we evaluate the performance of the Wide Field Survey Telescope (WFST) in identifying kilonova signals in composite afterglow-kilonova transients. Using a numerical framework based on the Fisher information matrix, we simulate $10,000$ realizations for each of two scenarios: an AT2017gfo-based template model and a physically sampled population that accounts for kilonova diversity. Our results indicate that kilonova identification is primarily limited by source distance. In both scenarios, the identification efficiency is largely insensitive to variations in afterglow microphysical parameters and exceeds $80\%$ at distances within approximately $600~\rm Mpc$ for AT2017gfo-like events. Under our adopted assumptions and a short gamma-ray burst (sGRB)-triggered target-of-opportunity (ToO) observational strategy, we estimate that the WFST could identify $0.1-1.2$ kilonovae per year in the optimistic scenario. Furthermore, we find that the discriminating power of color-based filters rapidly saturates, reaching a stable plateau by the second night after the merger. We therefore propose a staged observing strategy that prioritizes high-cadence $g$ and $r$-band monitoring during the first night and incorporates the $z$ band from the second night onward. This strategy improves the identification precision by exploiting the increasingly prominent red excess produced by the kilonova. Our results provide a physical basis for optimizing WFST observing resources to efficiently detect and characterize kilonovae in the multimessenger era.
\end{abstract}


\section{Introduction}\label{sec1}
The coalescence of binary neutron stars (BNS) systems represent an important class of multimessenger astrophysical events, producing gravitational-wave (GW) radiation accompanied by rich electromagnetic (EM) emission spanning a broad range of wavelengths \citep{Metzger2017, 2017RPPh, geng2018electromagnetic, Nakar2020, Metzger2020}. The merger process typically results in a compact remnant — either a massive neutron star (NS) or a black hole — surrounded by an accretion disk. The subsequent relativistic jet launch triggers a short gamma-ray burst (sGRB) \citep{eichler1989nucleosynthesis}, and the interaction between the jet and the interstellar medium produces a non-thermal afterglow spanning from X-ray to radio wavelengths \citep{meszaros1997optical, wijers1997shocked, sari1998observed, sari1999jets, berger2005afterglow, fox2005afterglow, hjorth2005optical}. Simultaneously, the neutron-rich ejecta undergoes rapid neutron-capture ($r$-process) nucleosynthesis, synthesizing heavy nuclei whose radioactive decay powers a thermal transient known as kilonova \citep{li1998transient, kulkarni2005modeling, metzger2010electromagnetic, kasen2013opacities}. Specifically, the X-ray afterglow typically persists for just several days, while optical afterglow lasts from days to weeks \citep{metzger2012most, berger2010short, fong2011optical, van2010off}.

While GW triggers provide a revolutionary pathway for identifying kilonovae, GW170817 \citep{abbott2017gw170817} and its associated signal GRB170817A remain the only event with a universally confirmed EM counterpart, AT2017gfo \citep{abbott2017multi, coulter2017swope, soares2017electromagnetic, valenti2017discovery}. A primary obstacle to targeted follow-up is the vast sky localization uncertainty inherent in GW triggers. For instance, according to the GWTC-2 catalog \citep{abbott2021gwtc}, the BNS merger GW190425 was localized within an immense region of $10,000 \, \mathrm{deg}^2$. Furthermore, kilonovae are intrinsically rare, subluminous, and rapidly evolving \citep{Metzger:2010sy}, making their identification even more challenging. A prominent recent case is the follow-up of the mass-gap GW event S250206dm; despite achieving deep $5\sigma$ limiting magnitudes across $64\%$ of the localization region, no associated kilonova was detected, underscoring the formidable observational barriers in confirming these elusive counterparts \citep{liu2026illuminating}.

Beyond GW triggers, kilonovae can also be identified through the follow-up of GRBs. Numerous kilonova candidates have been reported in association with GRBs, including GRBs050709 \citep{jin2016macronova}, 070707 \citep{zhu2023afterglow}, 070809 \citep{jin2020kilonova}, 080503 \citep{zhou2023grb}, 130603B \citep{tanvir2013kilonova}, 160821B \citep{troja2019afterglow}, 211211A \citep{rastinejad2022kilonova, yang2022long}, and 230307A \citep{levan2024heavy, yang2024lanthanide, Liu2025}. The majority of these events are classified as sGRBs, with the exceptions of 211211A and 230307A. However, the identification of kilonovae within GRB follow-up campaigns faces a formidable challenge: the presence of luminous afterglows, which can significantly mask the underlying kilonova emission \citep{kumar2015physics, fong2015decade}. Robustly discriminating them from afterglow-dominated transients is a technical necessity and a scientific priority for optimizing limited observational resources.

Current and future optical survey facilities, including the Zwicky Transient Facility (ZTF; \citealt{bellm2019zwicky}), the Vera C. Rubin Observatory's Legacy Survey of Space and Time (LSST; \citealt{abell2009lsst}), the Multichannel Photometric Survey Telescope (Mephisto; \citealt{wang2024prospects}), the SiTian Projects (Sitian, \citealt{liu2021sitian}) and the Wide Field Survey Telescope (WFST; \citealt{wang2023science,Liu2023,Yu2024}), are poised to expand the current kilonova samples. This work evaluates the performance of the WFST for identifying kilonova signals within the strategy of sGRB-triggered target-of-opportunity (ToO) follow-up observations, focusing on isolating kilonova components from complex multi-messenger light curves complicated by afterglow emission. To this end, we develop a comprehensive identification framework to identify kilonovae and quantify the detection efficiency, and further establish an optimized observational strategy based on multi-band color screening.

\begin{figure*}[h!]
\centering
\includegraphics[width=0.6\textwidth]{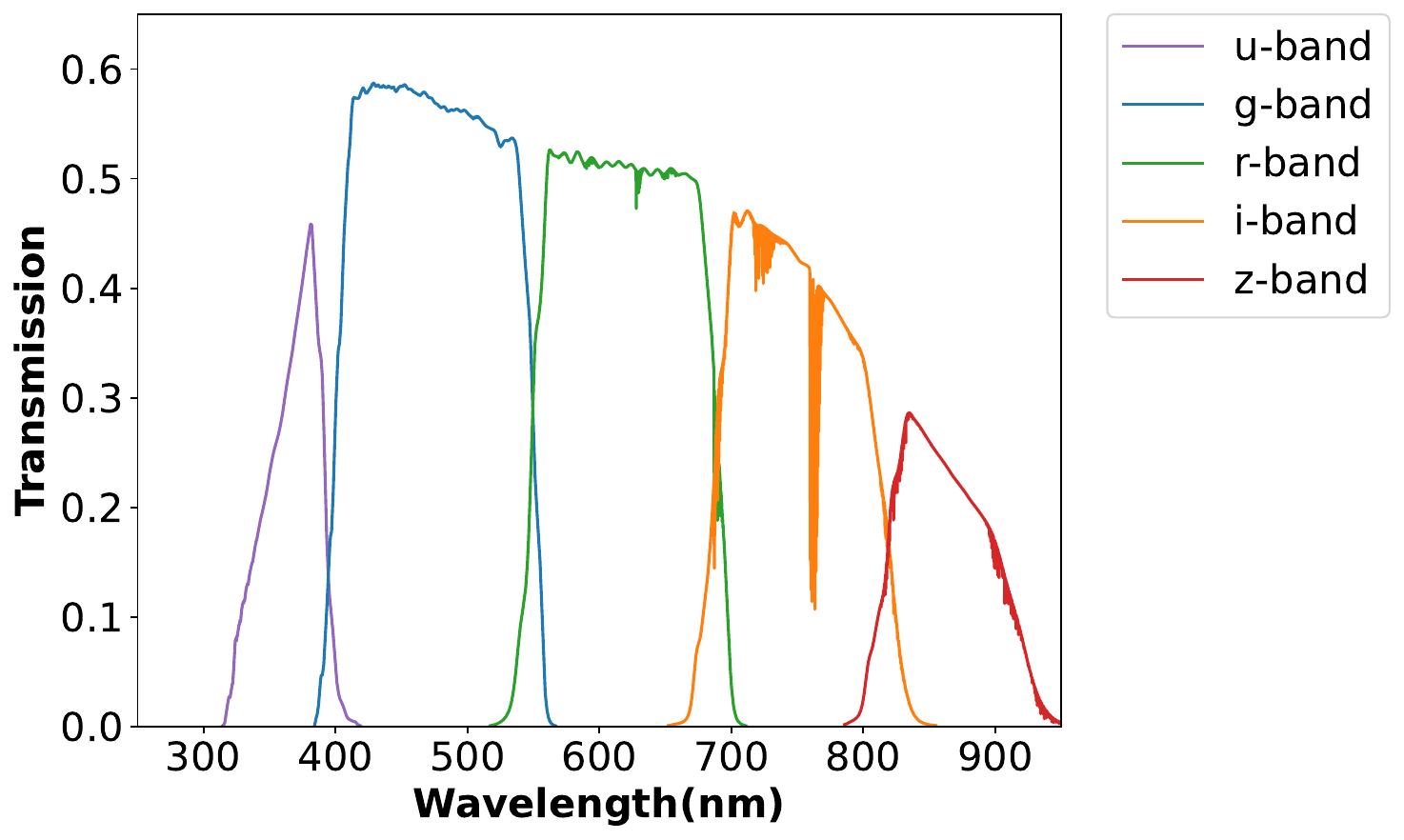}
\caption{The filter response curves of WFST bands ($u, g, r, i, z$). \label{fig:mutibands transmission}}
\end{figure*}

Located in Qinghai Province, China, at an altitude of $4,200 \, \mathrm{m}$, the WFST features a $2.5 \, \mathrm{m}$ primary mirror and a field of view optimized for high-cadence time-domain surveys. It covers a wavelength range of $320$--$1028 \, \mathrm{nm}$ across the $u, g, r, i, z,$ and $w$, surveying up to $6,000 \, \mathrm{deg}^2$ per night. For our analysis, we focus exclusively on the standard u,g,r,i, and z photometric bands. The filter response curves are illustrated in Figure~\ref{fig:mutibands transmission}, and the $5\sigma$ limiting magnitudes are summarized in Table~\ref{tab:limiting_magnitudes}, where the $5\sigma$ limiting magnitudes for a $30 \, \mathrm{s}$ exposure time are cited from \citep{wang2023science}. To evaluate the performance for deep-field observations of the combined afterglow and kilonova, we estimate the $5\sigma$ limiting magnitudes for a $300 \, \mathrm{s}$ exposure time based on the scaling relation of the signal-to-noise ratio $\mathrm{SNR} \propto \sqrt{F \cdot \tau}$, where $F$ is the signal flux, and $\tau$ denotes the exposure time. 
\begin{deluxetable*}{lcccccc}
\tablewidth{0pt}
\tablecaption{The $5\sigma$ limiting magnitudes of WFST bands for $30 \, \mathrm{s}$ exposure time and $300 \, \mathrm{s}$ exposure times (the former is adopted from \cite{wang2023science}). \label{tab:limiting_magnitudes}}
\tablehead{
\colhead{Magnitude} & \colhead{$u$} & \colhead{$g$} & \colhead{$r$} & \colhead{$i$} & \colhead{$z$} 
}
\startdata
Mag$_{30 \, \mathrm{s}}$ & 22.27 & 23.32 & 22.84 & 22.31 & 21.38  \\
Mag$_{300 \, \mathrm{s}}$ & 23.52 & 24.57 & 24.09 & 23.56 & 22.63  \\
\enddata
\end{deluxetable*}

This paper is structured as follows: In Section~\ref{sec:light curves}, we detail the numerical simulation of multimessenger light curves, incorporating both kilonova and afterglow components across a broad parameter space. Section~\ref{sec:identification framework} outlines our data processing methodology and the statistical identification framework based on the Fisher Information Matrix. In Section~\ref{sec:results}, we quantify the annual kilonova detection yield for the WFST under the sGRB-triggered ToO observing strategy and present the simulation results, contrasting a template-based model with a physically sampled BNS population to evaluate how kilonova diversity and afterglow environments influence identification performance. Section~\ref{sec:color} establishes our staged color-based filters and demonstrates the saturation of their efficiency by the second night post-merger. Finally, our conclusions and outlook for the WFST in the multi-messenger era are summarized in Section~\ref{sec:conclusion}.

\section{Generation of Multi-messenger Light Curves} \label{sec:light curves}
In this section, we describe the numerical framework used to generate the spectral energy distributions (SEDs) and synthetic multi-band light curves for both afterglow and kilonova components. For the afterglow, we employ \textsc{afterglowpy} \citep{ryan2020gamma}, a semi-analytical code for computing synchrotron radiation from relativistic jet interactions. For the kilonova, we utilize the \textsc{redback} package \citep{sarin2024redback} to interface with the \textsc{possis} model-a multi-dimensional Monte Carlo radiative transfer code that accounts for viewing-angle dependent observables \citep{bulla2019possis}. Subsequently, we use \textsc{sncosmo} \citep{barbary2016sncosmo} to convolve these flux densities with the WFST response functions, generating synthetic photometry.

\subsection{Afterglow Modeling} \label{AG}
The structured jet of an afterglow can be characterized by various profiles, including top-hat, Gaussian, and power-law models. Considering that the viewing angle of GRB170817A was approximately $20\,^\circ$ \citep{mooley2018mildly}, we adopt the Gaussian jet profile. This choice effectively captures scenarios involving a narrow jet core observed at relatively large inclinations. The model is defined by several key parameters: the viewing angle ($\theta_{\mathrm{obs}}$), the jet core half-opening angle ($\theta_c$), the outer truncation angle of the jet ($\theta_{\mathrm{trunc}}$), and the isotropic-equivalent energy along the jet axis ($E_0$). The circumburst environment is characterized by the proton number density ($n_0$), while the shock microphysics is governed by the electron energy distribution index ($p$) and the fractions of thermal energy in relativistic electrons ($\epsilon_e$) and the magnetic field ($\epsilon_B$). \citet{fong2015decade} presented a comprehensive catalog and analysis of broadband afterglow observations for 103 sGRBs over a decade, and \citet{becerra2023understanding} collected data from 227 GRBs observed by the TAROT, COATLI, and RATIR telescopes. We note the afterglow parameter degeneracies in their studies, where tightly coupled physical parameters can yield comparable observational light-curve fits. Therefore, we do not directly sample parameter values from their fitted results. Instead, we only draw upon their observational statistical outcomes to set physically motivated allowed ranges for our simulation parameters. Within our simulation framework, all relevant physical parameters are independent of one another. Based on their research, we assume uniform distributions for these parameters within the ranges specified in Table~\ref{tab:ag parameters}. For $E_0, n_0, \epsilon_e,$ and $\epsilon_B$, we adopt logarithmic scaling (i.e., $\log E_0, \log n_0, \log \epsilon_e, \log \epsilon_B$). Assuming the axes of BNS systems are isotropically distributed in space, we sample $\cos \theta_{\mathrm{obs}}$ uniformly. Given the strong dependence of afterglow luminosity on both the jet opening angle and the viewing angle, we limit the viewing angle to $\theta_{\mathrm{obs}} \leq 30\,^\circ$, consistent with previous studies (e.g., \citet{ryan2020gamma, perkins2026searching}). This threshold covers the full range of viewing angles, encompassing both on-axis and off-axis orientations in our simulations. The truncation angle is fixed at $\theta_{\mathrm{trunc}} = 3 \theta_c$ following \citet{gong2024off}. Regarding the remaining \textsc{afterglowpy} parameter, the fraction of accelerated electrons is fixed at its default value, $\xi_N = 1.0$.
\begin{deluxetable*}{llc}
\tablewidth{0pt}
\tablecaption{Afterglow parameters constrained by the results of \citet{fong2015decade} and \citet{becerra2023understanding}. \label{tab:ag parameters}}
\tablehead{
\colhead{Parameter} & \colhead{Meaning} & \colhead{Distribution}
}
\startdata
$\theta_{\mathrm{obs}}$ & Viewing angle ($^\circ$) & $\arccos(\mathcal{U}(\cos30\,^\circ, 1))$ \\
$\log E_0$ & Logarithm of isotropic energy ($\mathrm{erg}$) & $\mathcal{U}(49, 52)$ \\
$\log n_0$ & Logarithm of proton density ($\mathrm{cm}^{-3}$) & $\mathcal{U}(-4, -1)$ \\
$p$ & Electron energy distribution index & $\mathcal{U}(2.1, 2.6)$ \\
$\theta_c$ & Jet core half-opening angle (rad) & $\mathcal{U}(0.1, 0.4)$ \\
$\log \epsilon_e$ & Logarithm of electron energy fraction & $\mathcal{U}(-3, -1)$ \\
$\log \epsilon_B$ & Logarithm of magnetic energy fraction & $\mathcal{U}(-5, -2)$ \\
\enddata
\end{deluxetable*}

\subsection{Kilonova Modeling} \label{subsec:KN}
Following the \textsc{possis} framework, the light curves of kilonovae from BNS mergers are characterized by a set of physical and geometric parameters. These include the dynamical ejecta mass, $m_{\mathrm{ej}}^{\mathrm{dyn}}$, which is spatially distributed according to an opening angle $\pm \Phi$ that separates the high-opacity, lanthanide-rich material concentrated near the equatorial plane from the lower-opacity component \citep{bulla2019possis}. Concurrently, the material released from the post-merger remnant is encapsulated by a separate component, where late-time, visco-neutrino-driven outflows from the debris disk constitute the disk wind ejecta mass, $m_{\mathrm{ej}}^{\mathrm{wind}}$ \citep{metzger2008time}. The dynamical ejecta spans velocities of $0.08c-0.3c$, while the disk wind ejecta covers $0.025c-0.08c$ \citep{dietrich2020multimessenger}. The viewing angle, $\theta_{\mathrm{obs}}$, also serves as a crucial parameter, introducing strong direction-dependent effects on the observed emission. $\theta_{\mathrm{obs}}$ describes the observer's line-of-sight relative to the binary orbital axis, with no predefined correlation with the ejecta masses; these parameters are sampled independently.
 
To generate a physically motivated sample of kilonovae, we sample $\Phi$ uniformly from $15\,^{\circ}$ to $75\,^{\circ}$ \citep{shah2024predictions, perkins2026searching}, assuming the same viewing angle as that of the afterglow. Due to the limited availability of observational data, direct sampling of the ejecta mass distribution remains challenging. To address this limitation, we instead sample the masses of the NS progenitors and estimate the resulting ejecta properties through analytical relations. In the formation of BNS systems, the two progenitor stars undergo successive collapses. The first-born NS is typically spun up to rapid rotation through accretion and mass transfer from the helium core of its companion, forming a recycled NS, while the second-born companion remains a slow NS, characterized by a significantly lower spin rate \citep{tauris2017formation}. We adopt the double Gaussian distribution $\{\mu_1, \sigma_1, \mu_2, \sigma_2, \alpha\} = \{1.34\,M_{\odot}, 0.02\,M_{\odot}, 1.47\,M_{\odot}, 0.15\,M_{\odot}, 0.68\}$ for the recycled NS and a uniform distribution of $[1.16\,M_{\odot}, 1.42\,M_{\odot}]$ for the slow NS, as detailed in \citet{farrow2019mass}.

By analyzing Tolman-Oppenheimer-Volkoff (TOV) integrations across a comprehensive range of nuclear equations of state (EOSs), \citet{de2018tidal} established a universal empirical formula for the effective tidal deformability $\tilde{\Lambda}$, which is widely used in recent studies \citep[e.g.,]{coughlin2019multimessenger, abbott2020gw190425, wang2023study}. The formula is given by:
\begin{equation}
\tilde{\Lambda} \simeq 800\left(\frac{R_{1.4}}{11.2\,\mathrm{km}} \frac{M_{\odot}}{\mathcal{M}}\right)^6,
\end{equation}
where $\mathcal{M}=(m_1 m_2)^{3/5} (m_1+m_2)^{-1/5}$ is the chirp mass and $R_{1.4}$ is the radius of a $1.4\,M_{\odot}$ NS. The binary (or effective) tidal deformability is determined by a specific combination of $\Lambda_1$ and $\Lambda_2$ \citep{flanagan2008constraining, wade2014systematic, favata2014systematic}:
\begin{equation}
\tilde{\Lambda} = \frac{16}{13} \frac{(m_1+12 m_2) m_1^4 \Lambda_1 + (m_2+12 m_1) m_2^4 \Lambda_2}{(m_1+m_2)^5},
\end{equation}
where $\Lambda_1$ and $\Lambda_2$ are the tidal deformabilities of the two neutron stars, respectively. These are related by:
\begin{equation}
\Lambda_1 / \Lambda_2 = (m_2 / m_1)^6.
\end{equation}
With the three relations above, we solve for each tidal deformability $\Lambda_i$. The NS compactness, $C_i$, is then derived from $\Lambda_i$ through a Quasi-Universal Relation (QUR):
\begin{equation}
C_i = 0.36 - 0.0355 \ln \Lambda_i + 0.000705 (\ln \Lambda_i)^2,
\end{equation}
which is accurate to $\sim 6.5\%$ across a variety of EOSs \citep{yagi2017approximate}. Finally, we calculate the dynamical ejecta mass following the fitting formula from \citet{coughlin2019multimessenger}:
\begin{equation}
\log m_{\mathrm{ej}}^{\mathrm{dyn}} = \left[ a \frac{(1-2 C_1) m_1}{C_1} + b m_2 \left(\frac{m_1}{m_2}\right)^n + \frac{d}{2} \right] + [1 \leftrightarrow 2],
\end{equation}
with constants $a=-0.0719, b=0.2116, d=-2.42$, and $n=-2.905$. The notation $[1 \leftrightarrow 2]$ denotes the repetition of the preceding terms with the binary indices interchanged (i.e., $m_1 \leftrightarrow m_2$). The fractional uncertainty of this fit is $36\%$. To obtain the mass of disk wind ejecta, we first estimate the remnant disc mass \citep{coughlin2019multimessenger, radice2018binary}:
\begin{equation}
\log m_{\mathrm{disc}} = \max \left( -3, a \left( 1 + b \tanh \left[ \frac{c - M_{\mathrm{tot}} / M_{\mathrm{thresh}}}{d} \right] \right) \right),
\end{equation}
where $M_{\mathrm{tot}}$ is the total mass of the BNS, with constants $a=-31.335, b=-0.9760, c=1.0474$, and $d=0.05957$. $M_{\mathrm{thresh}}$ is the mass threshold for prompt black hole collapse, parameterized as \citep{bauswein2013prompt}:
\begin{equation}
M_{\mathrm{thresh}} = \left( 2.38 - 3.606 \frac{M_{\mathrm{TOV}}}{R_{1.4}} \right) M_{\mathrm{TOV}},
\end{equation}
where $M_{\mathrm{TOV}}$ is the maximum TOV mass. The disk wind ejecta mass is then given by $m_{\mathrm{ej}}^{\mathrm{wind}} = \eta_{\mathrm{disc}} m_{\mathrm{disc}}$, accounting for the efficiency of unbinding $\eta_{\mathrm{disc}}$ due to viscous processes \citep{radice2018viscous, radice2018binary}. For simplicity, we adopt fixed BNS simulation parameters: $R_{1.4} = 11.06\,\mathrm{km}$ and $M_{\mathrm{TOV}} = 2.17\,M_{\odot}$, which are the median values obtained by fitting the AT2017gfo light curve \citep{nicholl2021tight}, and assume $\eta_{\mathrm{disc}} = 0.2$ \citep{siegel2017three, radice2018binary}.
 
\section{Kilonova Identification Framework} \label{sec:identification framework}
In this section, we present our observational modeling and detail the numerical framework developed to identify kilonova signals within observable composite transients with the WFST. While the presence of an afterglow can enhance the total luminosity and improve initial detectability \citep{perkins2026searching}, an overwhelmingly dominant afterglow component can mask the distinctive temporal and spectral signatures of the underlying kilonova.

\subsection{Observational Error Modeling} \label{subsec:observation}
Our observational strategy relies on rapid sGRB-triggered ToO observations. Upon receiving external triggers from global GRB detectors, the WFST is programmed to rapidly slew to the localization region of the burst and perform follow-up observations. To identify as many kilonovae as possible from composite transients, our numerical framework implements a dense multi-band observational cadence spanning six consecutive nights. We simulate a realistic scheduling baseline for the WFST by assuming a nightly visibility window of 8 hours, commencing at $0.1$ days post-merger:
\begin{enumerate}
    \item \textbf{Initial Phase (Nights 1-2):} During the first two nights, a high-cadence monitoring protocol is adopted, with each band sampled once every $4$ hours. This frequency of three observations per night per band is designed to capture the rapid early-time evolution of the transient.
    \item \textbf{Late Phase (Nights 3-6):} From the third night onward, the cadence transitions to once per night as the light curve evolution becomes more gradual, continuing until the sixth night post-merger.
\end{enumerate}
In general, simulation studies typically consider kilonovae with redshift $z < 0.2$ \citep{abbott2023population}. However, due to the enhanced detection depth provided by the inclusion of afterglow emission, we extend the luminosity distance range to $1,400\,\mathrm{Mpc}$, which corresponds to the critical distance for kilonova identification in our study. We adopt a uniform spatial distribution for simulated BNS mergers across the considered volume. The maximum distance corresponds to a relatively low redshift ($z < 0.3$), within which cosmological evolution of the merger density is negligible. We therefore do not account for changes in merger density with redshift in our simulations, and source sampling is purely uniform in physical space.

To replicate realistic observing conditions, we inject observational uncertainties into the theoretical light curves following the methodology described in \citet{liu2023preliminary}. The photometric error $\sigma_{\mathrm{ph}}$ is derived from the SNR using the approximate relation \citep{pozzetti1998high, bolzonella2000photometric}:
\begin{equation}
\sigma_{\mathrm{ph}} = 2.5 \log \left( 1 + \frac{1}{\mathrm{SNR}} \right).
\end{equation}
To account for instrumental limitations and ensure numerical stability during the fitting process, a systematic error floor of $\sigma_{\mathrm{sys}} = 0.02\,\mathrm{mag}$ is introduced \citep{cao2018testing}. The quadrature sum of these components yields the total photometric uncertainty: 
\begin{equation}
\sigma_{\mathrm{total}} = \sqrt{\sigma_{\mathrm{ph}}^2 + \sigma_{\mathrm{sys}}^2}.
\end{equation}
Finally, to produce the simulated data, a random perturbation is drawn from a Gaussian distribution, $\mathcal{N}(0, \sigma_{\mathrm{total}}^2)$, and added to the theoretical magnitude. This procedure ensures that each simulated observation reflects both the expected statistical fluctuations and the sensitivity limits of the survey.

We implement a rigorous filtering process based on the $5\sigma$ limiting magnitudes $\mathrm{Mag}_{300 \, \mathrm{s}}$ presented in Table~\ref{tab:limiting_magnitudes}. A transient signal is classified as observable only if it satisfies a sequential set of multi-epoch and multi-band criteria, designed to suppress false positives and ensure sufficient data quality for subsequent modeling:
\begin{enumerate}
    \item \textbf{Discovery Phase (Night 1):} The source must be detected in at least two independent photometric bands during at least one observation epoch.
    \item \textbf{Confirmation Phase (Night 2):} Similar to the first night, a persistent detection in at least two bands is required to confirm the transient's nature and mitigate contamination from short-lived astrophysical noise.
    \item \textbf{Monitoring Phase (Night 3):} The signal must remain detectable in at least one photometric band to provide the necessary temporal baseline for light curve characterization.
\end{enumerate}
Only candidate signals that fulfill this entire suite of criteria are promoted to the subsequent identification analysis. This ensures that the parameter estimation for kilonova identification is performed only on sources with high-fidelity data.
 
\subsection{Statistical Identification Method} \label{identification method}
The identification of a kilonova component within a complex multi-wavelength signal hinges on the reliable detection of excess flux relative to the underlying afterglow emission. Historically, such excesses have been pivotal in characterizing kilonova properties. For instance, the nIR (F160W) excess observed in GRB130603B provided evidence for an ejecta mass ranging from $10^{-2}\,M_{\odot}$ to $10^{-1}\,M_{\odot}$ \citep{tanvir2013kilonova}. Similarly, GRB160821B exhibited clear optical and nIR excesses similar to the AT2017gfo template \citep{troja2019afterglow}. Recent multi-wavelength modeling has also revealed sophisticated components, such as the engine-fed kilonova in GRB211211A \citep{yang2022long} and the two-component kilonova emission observed in the exceptionally bright GRB230307A \citep{yang2024lanthanide}. Motivated by these precedents, we employ the Fisher Information Matrix (FIM) \citep{fisher1925theory} to distinguish the kilonova-originated excess from the afterglow background.

The FIM formalism is a robust and widely adopted framework for parameter estimation and experimental design in diverse astronomical contexts \citep{tegmark1997karhunen, albrecht2006report}. In this study, we model the observed multi-wavelength flux density as a linear combination of afterglow and kilonova:
\begin{equation}\label{aaa}
f_{\nu, \mathrm{total}}(\mathbf{p}, A) = f_{\nu, \mathrm{afterglow}}(\mathbf{p}) + A \cdot f_{\nu, \mathrm{kilonova}},
\end{equation}
where $\mathbf{p}$ represents the vector of physical parameters governing the afterglow (e.g., $\log{E_0}, \log{n_0}, \log{\epsilon_e}$). In principle, the flux density $f_{\nu,\mathrm{kilonova}}$ should contain the corresponding kilonova parameters as discussed in previous section. However, the well constraints on all these parameters seems impossible, based on the relatively weak signal of kilonova. In practice, in most cases of real observation, the detection of kilonova is always defined by the deviation of observed flux density from the pure afterglow model, rather than the well constraints on the kilonova parameters. Therefore, as an optimistic and simple estimation, in this article we quantify the contribution of kilonova component by a simple parameter $A$ in Eq.~(\ref{aaa}) \footnote{Similar approach is often used in cosmological research as well, such as the detection of cosmic weak lensing in cosmic microwave background radiation \citep{planck}.}, which denotes the dimensionless amplitude of the kilonova template. By definition, $A=1$ corresponds to a physical match with the theoretical model. Then we convert the flux density into the corresponding magnitude $m_{\mathrm{total}}$. To quantify the precision of parameter recovery, we compute the elements of the Fisher matrix $F_{ij}$ by evaluating the sensitivity of the total model to variations in the parameter space. For any parameter $p_i$ (including $A$), the partial derivatives are estimated via the central finite difference method:
\begin{equation}
\frac{\partial m_{\mathrm{total}}}{\partial p_i} \approx \frac{m_{\mathrm{total}}(p_i + \epsilon_i) - m_{\mathrm{total}}(p_i - \epsilon_i)}{2\epsilon_i},
\end{equation}
where $\epsilon_i$ is a sufficiently small perturbation ensuring numerical stability. The Fisher matrix is then constructed by summing the weighted gradients over all observable epochs and bands:
\begin{equation}
F_{ij} = \sum_{k} \frac{1}{\sigma_{k}^2} \frac{\partial m_{\mathrm{total}, k}}{\partial p_i} \frac{\partial m_{\mathrm{total}, k}}{\partial p_j},
\end{equation}
where $\sigma_{k}$ is the total error, which represents the combined observational and statistical uncertainties at the $k$-th data point. According to the Cramér-Rao Lower Bound (CRLB) \citep{kay1993statistical}, the covariance matrix $\mathbf{C}$ is given by the inverse of the Fisher matrix. The marginal uncertainty $\sigma_A$ on the kilonova amplitude, is derived from the corresponding diagonal element:
\begin{equation}
\sigma_{A} = \sqrt{(F^{-1})_{AA}}.
\end{equation}
In this framework, we require a $5\sigma$ statistical confidence for the identification of the kilonova component, defined by the criterion $\sigma_{A} \leq 0.2$. This threshold ensures that the kilonova can be reliably distinguished from the afterglow background with high statistical significance.

\begin{figure*}[!ht]
\centering
\includegraphics[width=1\textwidth]{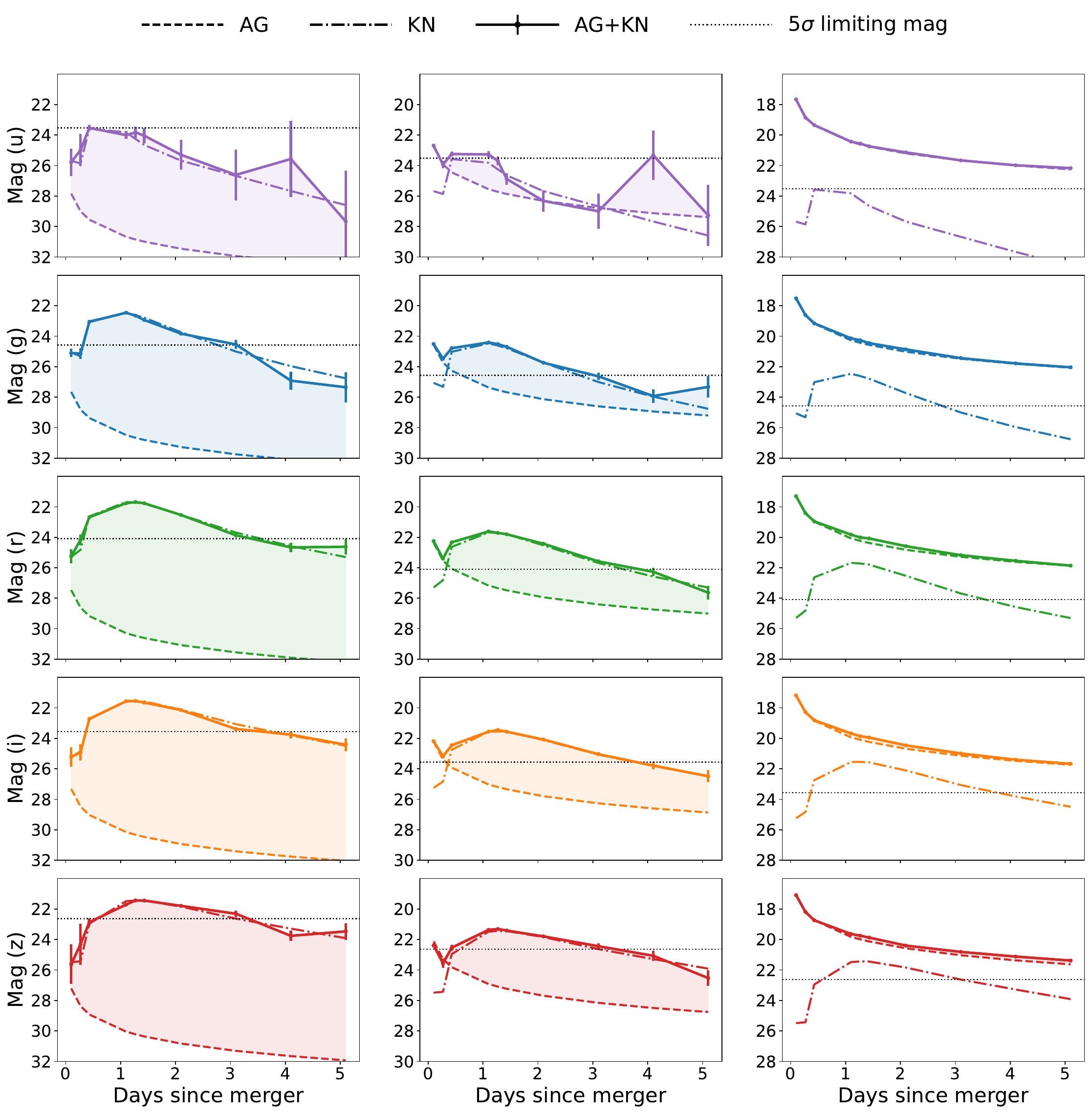}
\caption{Multi-band light curves of three representative simulated samples. In each band ($u, g, r, i, z$), the solid lines with error bars denote the total observed magnitudes and black dotted lines represent the $5\sigma$ limiting magnitudes. Theoretical components are decomposed into the afterglow (dashed lines) and the kilonova (dash-dotted lines). The shaded regions highlight the flux excess contributed by the kilonova relative to the afterglow background. All samples are placed at a fixed distance of $400\,\mathrm{Mpc}$. The kilonova parameters are fixed to the best-fit values of the AT2017gfo template: $m_{\mathrm{ej}}^{\mathrm{dyn}} = 10^{-2.27}\,M_{\odot}$, $m_{\mathrm{ej}}^{\mathrm{wind}} = 10^{-1.28}\,M_{\odot}$, $\Phi = 49.5\,^\circ$ \citep{dietrich2020multimessenger}. The afterglow parameters are set to the median values of our adopted distributions: $\theta_{\mathrm{obs}}=21.09\,^\circ$, $\log{n_0}=-2.5$, $p=2.35$, $\theta_c=0.25$, $\log{\epsilon_e}=-2.0$, $\log{\epsilon_B}=-3.5$, while $\log{E_0}= 49.0,50.5,52.0$ from left to right, respectively. \label{fig:lightcurve samples}} 
\end{figure*}

Figure~\ref{fig:lightcurve samples} illustrates the efficacy of this framework. In the left and middle panels, the emergence of the kilonova component induces a significant deviation from the standard afterglow decay, especially in the redder bands ($g,r,i,z$). This pronounced chromatic signature provides the requisite information for our algorithm to successfully distinguish the kilonova signal from the afterglow background, yielding robust identification results ($\sigma_{A}=0.111$ and $0.088$, respectively). Conversely, the right panel illustrates a scenario where the kilonova component remains unidentified. In this case, the total emission is overwhelmingly dominated by the bright afterglow background across all filters. The kilonova signal is deeply masked, resulting in an observed light curve that is nearly indistinguishable from a pure afterglow. Consequently, our framework identifies this event as unconstrained ($\sigma_{A}=0.222$), demonstrating its reliability in avoiding false-positive kilonova identifications when the kilonova component is completely masked.

\section{Results of Parameter Distribution}\label{sec:results}
To evaluate the identification performance of the WFST, we conduct Monte Carlo simulations consisting of $10,000$ realizations for each scenario. \textbf{Case 1: Template-Based Kilonova.} This case combines the afterglow model (parameters detailed in Table~\ref{tab:ag parameters}) with a kilonova component fixed to the best-fit parameters of AT2017gfo: $m_{\mathrm{ej}}^{\mathrm{dyn}} = 10^{-2.27}\,M_{\odot}$, $m_{\mathrm{ej}}^{\mathrm{wind}} = 10^{-1.28}\,M_{\odot}$, and $\Phi = 49.5\,^\circ$ \citep{dietrich2020multimessenger}. \textbf{Case 2: Physically Sampled Kilonova.} In this scenario, while the afterglow parameters remain consistent with Table~\ref{tab:ag parameters}, the kilonova properties are stochastically derived from the physical parameter space described in Section~\ref{subsec:KN}.

\subsection{\textbf{Case 1: Template-Based Kilonova}}\label{subsec:Case 1}
Within the $n_{\mathrm{total}}=10,000$ simulated samples in Case 1, we identify $n_{\mathrm{KN}}=1,449$ samples where the kilonova component is successfully recovered with $5\sigma$ significance ($\sigma_{A} \leq 0.2$). To project these simulation results into an annual identification rate for the WFST, we adopt the most recent volumetric BNS merger rate estimate, $R=56^{+99}_{-40}\,\mathrm{Gpc}^{-3}\,\mathrm{yr}^{-1}$, derived from the Gravitational-Wave Candidate Event Database alert stream \citep{akyuz2025mining}. The annual number of BNS mergers occurring within our simulated cosmic volume is
\begin{equation}
N_{\mathrm{BNS}} = R \cdot V_{\mathrm{max}}=643.7^{+1138}_{-459.8}\ \mathrm{yr}^{-1},
\end{equation}
where $V_{\mathrm{max}}$ is the volume corresponding to our distance limit of $1,400\,\mathrm{Mpc}$. Fermi-GBM had detected 395 sGRBs over the 2008-2018 period, corresponding to an annual detection rate of $39.5~\mathrm{yr}^{-1}$ \citep{von2020fourth}. The observed sGRB detection rate is modulated by a variety of astrophysical and instrumental selection effects. sGRBs arise from collimated relativistic outflows, and are only detectable when the line of sight falls within the jet opening angle. For our adopted viewing angle threshold, the geometric probability that a randomly oriented jet points toward Earth is
\begin{equation}
f_{\theta \leq 30\,^\circ} = \frac{\int_{0}^{30\,^\circ} \sin\theta \,\mathrm{d}\theta}{\int_{0}^{90\,^\circ} \sin\theta \,\mathrm{d}\theta}.
\end{equation}
However, not all geometrically visible jets produce prompt emission bright enough to trigger Fermi-GBM, due to the intrinsic dispersion in jet kinetic energies and the angular dimming of the beamed emission. To address this, we generate 10,000 sGRB jets by sampling four physical parameters (luminosity distance, the viewing angle, isotropic-equivalent kinetic energy, and jet core half-opening angle), according to the distributions specified in Table \ref{tab:ag parameters}. For each simulated jet, we compute the isotropic-equivalent kinetic energy at the viewing angle from the axial reference energy corresponding to the adopted Gaussian jet profile:
\begin{equation}
E_{\rm iso}(\theta_{\rm obs}) = E_0 \exp\left( -\frac{\theta_{\rm obs}^2}{2\theta_c^2} \right),
\end{equation}
We then scale this kinetic energy to prompt gamma-ray energy using a $\gamma-$ray efficiency $\eta_\gamma$. The full bolometric (1~keV-10~MeV) radiative efficiency of sGRBs is constrained to 15\% by multi-wavelength afterglow modeling \citep{beniamini2015energies}. Since the 50-300~keV Fermi-GBM trigger band captures only 1/10 to 1/3 of the total bolometric prompt emission \citep{poolakkil2021fermi}, we adopt an equivalent conversion factor of $\eta_\gamma = 3\%$ for scaling jet kinetic energy to gamma-ray energy in the GBM energy band. We derive the 50-300~keV peak photon flux at the corresponding luminosity distance. We adopt a characteristic prompt emission duration of $T_{90} \approx 0.25 \rm~s$ for the sGRB samples, consistent with the observed peak of the short-duration class in the well-established bimodal $T_{90}$ distribution \citep{kouveliotou1993identification},and a characteristic mean photon energy of $\langle E_{\rm ph} \rangle = 100\ \mathrm{keV}=1.6\times10^{-7}\ \mathrm{erg}$. The peak photon flux is then given by:
\begin{equation}
F_{\rm ph} = \frac{\eta_\gamma E_{\rm iso}(\theta_{\rm obs})}{4\pi d_L^2 T_{90} \langle E_{\rm ph} \rangle}.
\end{equation}
We apply the standard on-board trigger threshold of Fermi-GBM: $0.74\ \mathrm{ph\ cm^{-2}\ s^{-1}}$ in the 50-300~keV energy band \citep{meegan2009fermi}. We find that 4,236 sGRB samples are detectable within our 1400~Mpc survey volume, corresponding to 42.36\% of geometrically visible events $\theta_{\rm obs} \leq 30^\circ$ within our 1400~Mpc survey volume satisfy the flux detection criterion. We denote this flux detection efficiency as $f_{\mathrm{flux}} \approx 0.4236$.

Furthermore, not all BNS mergers successfully launch a relativistic jet that penetrates the merger ejecta and powers a detectable sGRB. The jet production efficiency $f_{\mathrm{prod}}$ is constrained to $0.69_{-0.37}^{+0.27}$ at 90\% confidence \citep{sarin2022linking}; we adopt the median value $f_{\mathrm{prod}}=0.69$ in our calculations. While individual GRB detectors have limited instantaneous FOV due to Earth occultation, the InterPlanetary Network (IPN) combines instruments in geocentric and interplanetary orbits to achieve nearly all-sky coverage. Importantly, the IPN only improves the localization accuracy of already detected bursts and does not increase the total number of detected events. We adopt Fermi-GBM detections as our primary sGRB trigger sample, consistent with previous studies \citep{troja2017x, mong2021searching, ahumada2022search}. Fermi-GBM has an instantaneous field-of-view (FOV) of 8.5 steradians \citep{meegan2009fermi}, corresponding to a sky coverage fraction $f_{\mathrm{FOV}} \approx 0.676$. We further apply an effective livetime fraction $f_{\mathrm{live}} \approx 0.85$ for Fermi-GBM, accounting for detector high-voltage shutdowns during South Atlantic Anomaly passages, data transmission gaps, and other instrumental deadtime \citep{goldstein2017ordinary}. Combining all the above selection effects, the annual rate of sGRBs within our simulated cosmic volume is
\begin{equation}
N_{\mathrm{sGRB}} = N_{\mathrm{BNS}} \cdot f_{\mathrm{prod}} \cdot f_{\theta \leq 30\,^\circ} \cdot f_{\mathrm{FOV}} \cdot f_{\mathrm{live}} \cdot f_{\mathrm{flux}}= 14.5_{-10.3}^{+25.6}~\mathrm{yr}^{-1}.
\end{equation}
We note that the small number of spectroscopically confirmed low-redshift sGRBs in the Fermi-GBM catalog does not necessarily imply a true deficit of nearby events. In the 10-year Fermi-GBM spectral catalog presented by \citet{poolakkil2021fermi}, which includes a total of 2,297 GRB events, only approximately 130 objects have spectroscopically confirmed redshift measurements, and merely 13 are classified as sGRBs, reflecting the very low success rate of redshift measurements for sGRB population. This discrepancy is probably attributable to the coarse localization of Fermi-GBM detections, which hinders follow-up observations to improve spectroscopic redshift measurements.

The localization coverage correction factor $f_{\mathrm{loc}}$ is a critical selection effect in ToO detection estimates, which quantifies the probability that the true burst position falls within the telescope's pointing footprint. As formulated in \citet{zhu2023kilonovae} for GW triggered follow-ups, this factor is derived from the geometric overlap between the telescope FOV and the 90\% confidence localization region of the trigger. We extend this standard framework to the case of Fermi-GBM sGRB triggers, whose localization uncertainties are considerably larger than the $3 ^\circ$ instantaneous FOV of the WFST. We adopt a nominal 90\% confidence localization radius of $R_{90} = 5^\circ$ for bright sGRBs that trigger ground-based follow-up \citep{connaughton2015localization}. To cover the error region efficiently, we employ a standard $3\times3$ mosaic pointing strategy for ToO observations, which consists of 9 individual pointings spaced by one WFST FoV diameter. The area of the 90\% confidence error circle is $\Omega_{\mathrm{err}} = \pi R_{90}^2$. Four circular segments of the error circle extend beyond the mosaic boundary. Its area is given by
\begin{equation}
\Omega_{\mathrm{seg}} = R_{90}^2 \arccos(\frac{4.5}{R_{90}}) - 4.5\sqrt{R_{90}^2 - 4.5^2}.
\end{equation}
The total overlapping area between the error circle and the mosaic is $\Omega_{\mathrm{overlap}} = \Omega_{\mathrm{err}} - 4\Omega_{\mathrm{seg}}$, corresponding to a geometric coverage fraction:
\begin{equation}
P_{\mathrm{geo}} = \frac{\Omega_{\mathrm{overlap}}}{\Omega_{\mathrm{err}}}.
\end{equation}
Real WFST pointings have an effective usable area of only $6.5\ \mathrm{deg^2}$ within the $9.0\ \mathrm{deg^2}$ footprint due to CCD gaps and truncated corners \citep{wang2023science}. Combined with the intrinsic 90\% confidence of the GBM localization, the estimated geometric coverage probability is $f_{\mathrm{loc}}=0.9 \cdot P_{\mathrm{geo}} \cdot \frac{6.5}{9.0} \approx 0.6$.

The WFST is sited to observe the northern sky, yielding an effective sky coverage fraction of $f_{\mathrm{sky}}=0.5$. The actual nightly astronomical observing window varies strongly with season, ranging from $5~\mathrm{h}$ in June to more than $11~\mathrm{h}$ in January, with a mean effective observing time of $8~\mathrm{h}$ and only 70\% of nights are photometrically clear \citep{wang2023science}. We therefore introduce a simplified correction factor $f_{\mathrm{clear}}=0.7$. Our estimation of the kilonova identification rate follows the standard simulation-based framework with selection effect corrections, which is widely adopted in time-domain transient surveys \citep{scolnic2017complete}. The expected annual identified count $N_{\mathrm{id}}$ is given by:
\begin{equation}
N_{\mathrm{id}} = \frac{n_{\mathrm{KN}}}{n_{\mathrm{total}}} \cdot N_{\mathrm{sGRB}} \cdot f_{\mathrm{loc}} \cdot f_{\mathrm{sky}} \cdot f_{\mathrm{clear}}.
\end{equation}
Combining all these correction factors, we estimate that the WFST can identify $N_{\mathrm{id}} = 0.4^{+0.8}_{-0.3}$ kilonovae per year embedded in composite composite afterglow-kilonova signals. Nevertheless, our selection effect treatment adopts a simplified analytical approach. Additional losses from lunar skyglow, high-airmass observations, and other observational systematics are not included, which would further reduce the actual detection yield. A more comprehensive performance assessment would require dedicated injection-recovery simulations similar to those of \citet{Andreoni2021}. Our derived identification rates should therefore be interpreted as optimistic analytical estimates.

Additionally, our baseline simulations adopt an optimal timing assumption: observations start at $0.1~\mathrm{d}$ post-merger, under the premise that the event occurs precisely at the onset of a nighttime observing window. In reality, sGRBs and their associated kilonovae occur at arbitrary times throughout the day. The delay between the merger epoch and the first WFST observation is therefore governed by the diurnal phase of the trigger relative to the telescope's fixed nighttime observing window. For events occurring during daytime hours, observations cannot commence until the next available nighttime window, corresponding to a first-observation delay ranging from $0.1~\mathrm{d}$ up to $0.77~\mathrm{d}$. To verify this effect, we perform additional simulations with three different initial observation epochs: $t_0=0.2~\mathrm{d}$, $0.5~\mathrm{d}$, and $0.8~\mathrm{d}$ for extreme scenario. Each simulation contains 10,000 samples and adopts the same observational strategy and identification pipeline as the fiducial setup. We find that the kilonova component is successfully recovered in 1,416 samples for $t_0=0.2~\mathrm{d}$, 1,128 samples for $t_0=0.5~\mathrm{d}$, and 806 samples for $t_0=0.8~\mathrm{d}$. The moderate reduction in identification with increasing delay is physically expected, since afterglows are decreasing with time, which affects the luminosity of the combined signals. But kilonova emission peaks at ~1-2 days and our identification strategy relies dominantly on post-peak photometric features, a moderate delay only removes the faint early rising segment and does not substantially degrade the discriminative power. Even for the extreme scenario of $0.8~\mathrm{d}$, more than half of the detectable kilonovae can still be reliably identified, so our overall results remain valid.

\begin{figure}[h!]
  \begin{subfigure}[h]{1\textwidth}
    \centering
    \includegraphics[width=\textwidth]{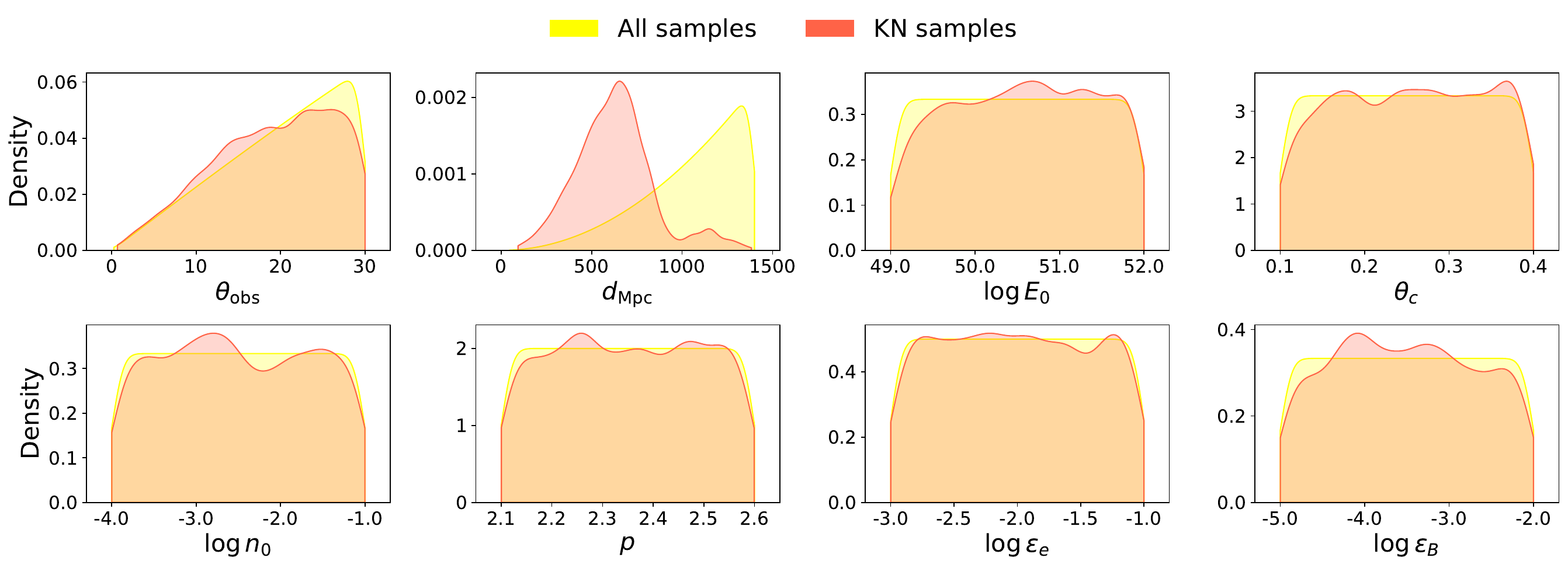}
    \caption{Parameter distributions \label{fig:a of 170817}}
  \end{subfigure}  
  \par\medskip
  \begin{subfigure}[h]{1\textwidth}
    \centering
    \includegraphics[width=\textwidth]{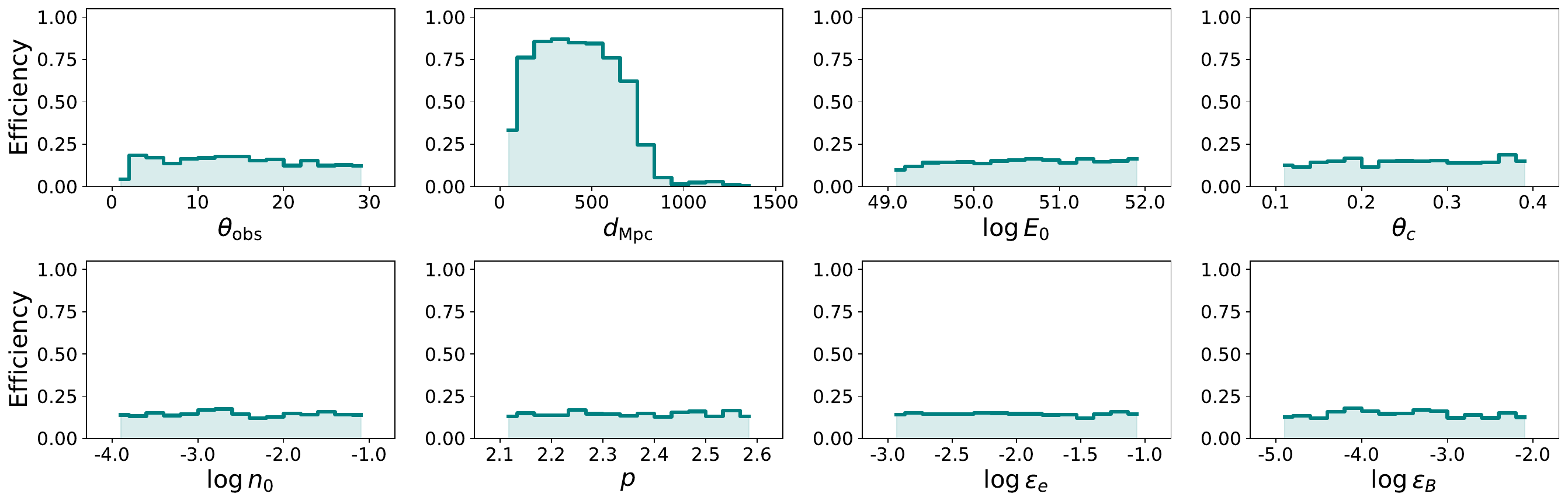}
    \caption{Identification efficiency \label{fig:b of 170817}}
  \end{subfigure}
  \caption{Statistical analysis of afterglow parameters for Case 1. \textbf{(a):} Kernel Density Estimation distributions of afterglow parameters for the total simulated samples (yellow) and the kilonova-identified samples (red). \textbf{(b):} Identification efficiency across the parameter space. Parameters include: viewing angle $\theta_{\mathrm{obs}}$, distance $d_{\mathrm{Mpc}}$, jet energy $\log E_0$, half jet angle $\theta_c$, circumburst density $\log n_0$, electron index $p$, and energy fractions $\log \epsilon_e$ and $\log \epsilon_B$. \label{fig:one-dim distribution 170817}} 
\end{figure}

To evaluate the identification performance of the WFST and its sensitivity to the physical environment of the afterglow, we analyze the parameter distributions of the identified sub-population $n_{\mathrm{KN}}$ relative to the total ensemble $n_{\mathrm{total}}$. While Figure~\ref{fig:a of 170817} illustrates the comparison between the parameter distributions of the identified samples and total samples, these primarily represent the observational selection effects of the survey, showing a clear preference for events at closer distances (approximately $600\,\mathrm{Mpc}$) and larger viewing angles (approximately $25\,^\circ$). However, to precisely determine which afterglow parameters facilitate kilonova identification, one must refer to the intrinsic identification efficiency shown in Figure~\ref{fig:b of 170817}, calculated as the ratio of sample counts within each bin, $\Delta n_{\mathrm{KN}}/\Delta n_{\mathrm{total}}$, effectively acting as a local normalization that eliminates the bias of the underlying distribution. Our findings indicate that the identification capability of the WFST is remarkably robust and is primarily governed by source distance rather than specific afterglow configurations. In the distance dimension, the identification efficiency remains above $80\%$ within approximately $600\,\mathrm{Mpc}$ but drops precipitously as distance increases, confirming that signal detectability is the primary constraint. Notably, the efficiency curves remain remarkably flat across a broad range of parameters, including jet energy ($\log E_0$) and circumburst density ($\log n_0$). This demonstrates that kilonova identification is robust against afterglow diversity, whether the afterglow is driven by a high-energy jet or situated in an extremely tenuous or dense interstellar medium. Consequently, the identification window for the WFST is essentially distance-limited rather than parameter-limited.

\subsection{\textbf{Case 2: Physically Sampled Kilonova}}\label{subsec:Case 2}
\begin{figure*}[h!]
\centering
\includegraphics[width=1\textwidth]{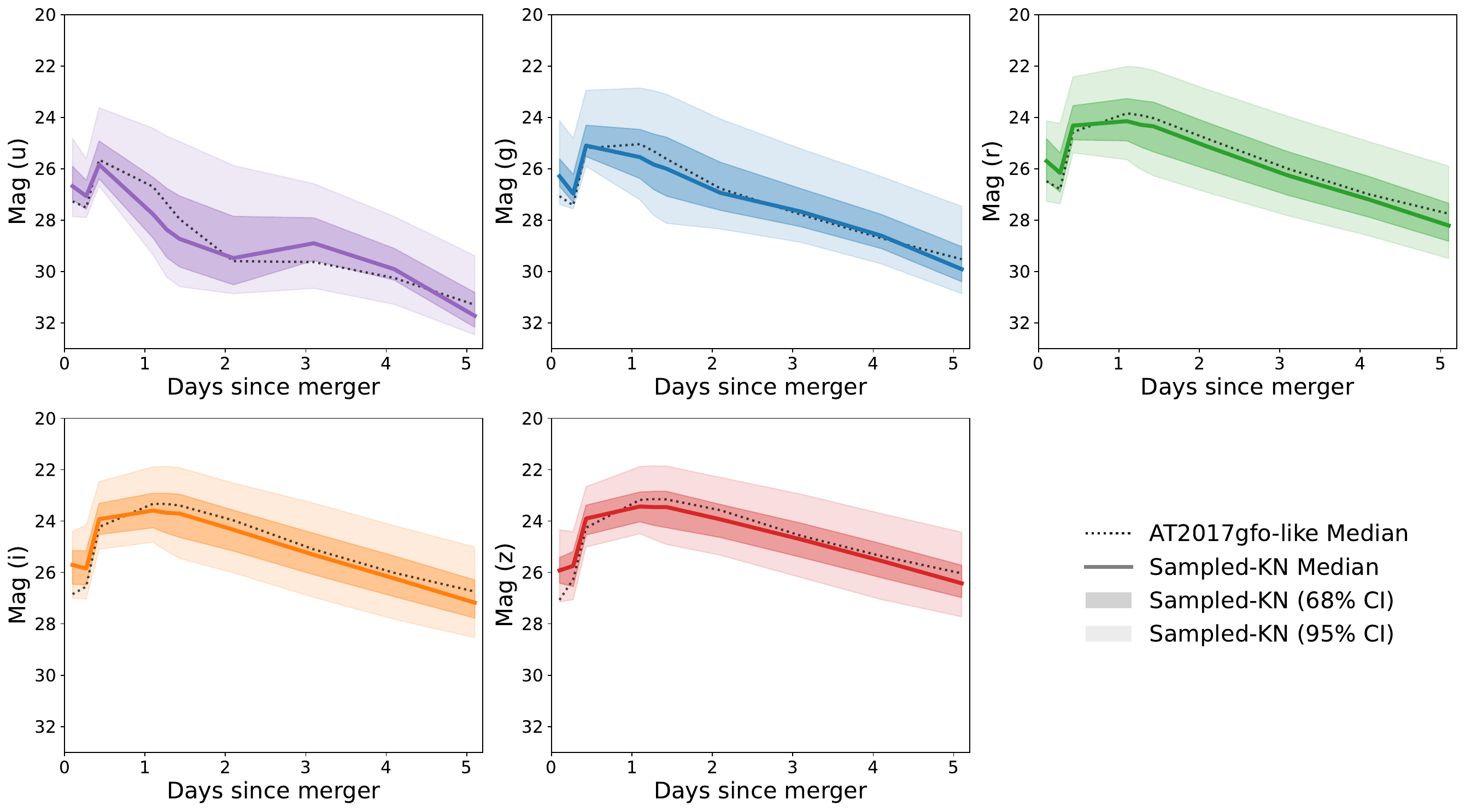}
\caption{Comparison of multi-band ($u, g, r, i, z$) light curves between AT2017gfo-like and sampled kilonova populations. The solid colored lines represent the median light curves of the sampled kilonova ensemble, with shaded regions indicating the $68\%$ and $95\%$ confidence intervals (CI). The black dotted line denotes the median light curves of the AT2017gfo-like samples. \label{fig:knmagcomparison}}
\end{figure*}

In Case 2, we identify $n_{\mathrm{KN}}=934$ kilonova samples, representing a reduction of approximately $35\%$ compared to Case 1. The corresponding estimated annual identified count is $N_{\mathrm{id}}=0.3^{+0.5}_{-0.2}$. This discrepancy primarily occurs because AT2017gfo represents an intrinsically luminous kilonova. As illustrated in Figure~\ref{fig:knmagcomparison}, the median magnitudes of the AT2017gfo-like model consistently resides near or above the median of the physically sampled population across all bands ($u, g, r, i, z$), particularly during the peak emission phase. 

\begin{figure}[h!]
  \begin{subfigure}[h]{1\textwidth}
    \centering
    \includegraphics[width=\textwidth]{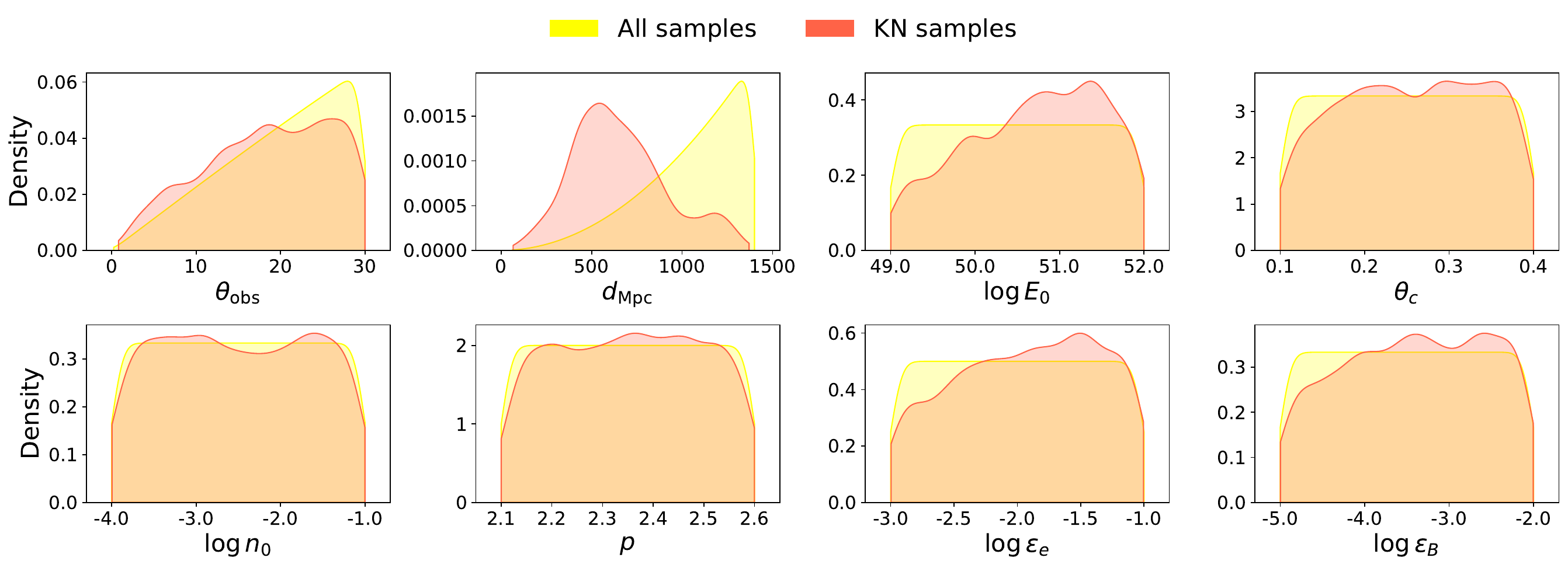}
    \caption{Parameter distributions \label{fig:a of kilo}}
  \end{subfigure}  
  \par\medskip
  \begin{subfigure}[h]{1\textwidth}
    \centering
    \includegraphics[width=\textwidth]{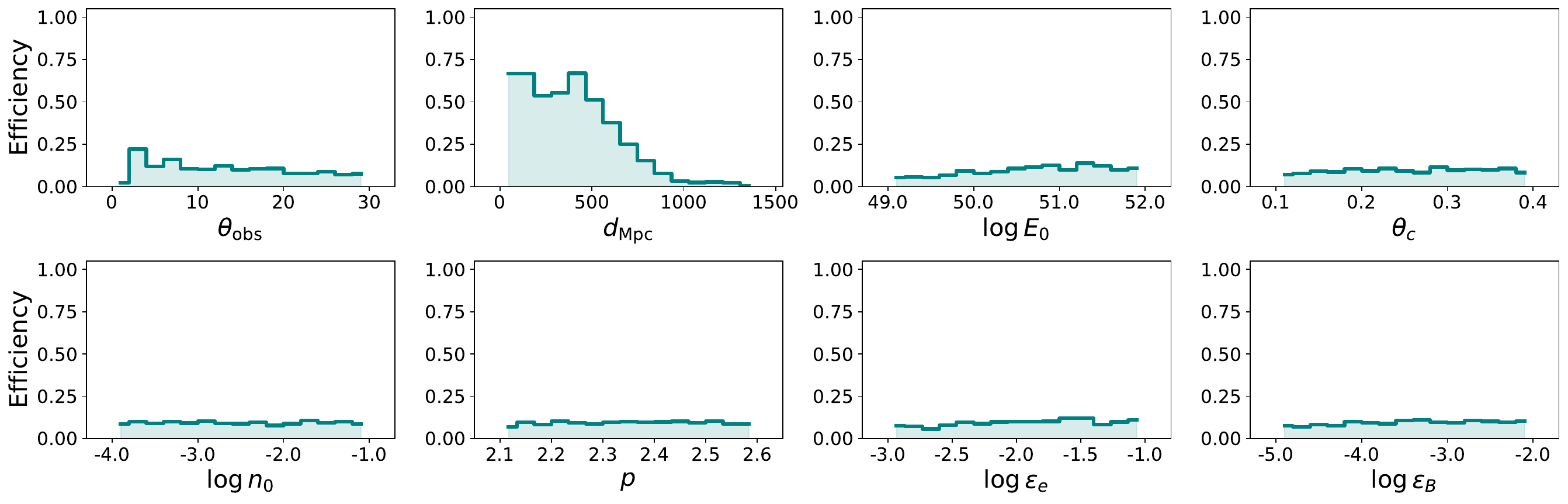}
    \caption{Identification efficiency \label{fig:b of kilo}}
  \end{subfigure}
  \caption{Statistical analysis of afterglow parameters for Case 2. \textbf{(a):} Kernel Density Estimation distributions of afterglow parameters for the total simulated samples (yellow) and the kilonova-identified samples (red). \textbf{(b):} Identification efficiency across the parameter space. Parameters include: viewing angle $\theta_{\mathrm{obs}}$, distance $d_{\mathrm{Mpc}}$, jet energy $\log E_0$, half jet angle $\theta_c$, circumburst density $\log n_0$, electron index $p$, and energy fractions $\log \epsilon_e$ and $\log \epsilon_B$. \label{fig:one-dim distribution kilo}}
\end{figure}

The parameter distributions for the randomized population, shown in Figure~\ref{fig:a of kilo}, are qualitatively consistent with the template-based results, reinforcing the dominance of distance and viewing angle as primary selection filters. However, a divergence is observed in the $\log E_0$ distribution, which exhibits a subtle shift toward higher energies. This trend suggests that for the diverse kilonovae in Case 2, a brighter afterglow background is required to boost the total luminosity. Compared to the AT2017gfo-like ensemble, the overall identification efficiency for the randomized population is systematically lower in Figure~\ref{fig:b of kilo}, particularly at intermediate distances. This reduction stems from the inclusion of a wider range of parameters within our model, which encompasses intrinsically fainter candidates. Despite this lower baseline, the efficiency curves across the afterglow microphysical parameters remain flat, confirming that the robustness against varying afterglow environments is a universal feature, independent of the specific kilonova parameter used. Importantly, although our simulation adopts a fixed distribution for the afterglow parameters, the resulting kilonova identification efficiency remains entirely decoupled from these microphysical variations and depends solely on distance. This independence, in turn, reinforces the intrinsic reliability and objectivity of our findings.

\section{Color-Based Identification Framework}\label{sec:color}
Kilonovae are characterized by their distinctively red spectra, a consequence of r-process nucleosynthesis. This property distinguishes them from afterglow emission, which is typically bluer due to its synchrotron origin. Color analysis targeting the red excess serves as a key diagnostic to characterize distinct kilonova components. For example, such color analysis was employed for the Vera C. Rubin Observatory to optimize kilonova identification \citep{stevenson2026strategy}. In this section, we investigate the color characteristics of the simulated transients to establish an effective filter for kilonova identification. We utilize the $10,000$-sample sets for both the Template-Based Kilonova and Physically Sampled Kilonova generated in section~\ref{sec:results}, and focus on the observable subset with the WFST. 

In practical time-domain surveys, allocating subsequent multi-band follow-up observations for every transient candidate is heavily constrained by telescope limited resource. Therefore, our primary objective is to implement these color filters as a sequential screening mechanism to isolate potential kilonova candidates from minimal early-time data. By establishing robust color cuts, we can introduce a decisive process to optimize resource allocation: candidates exhibiting a clear red excess are strategically prioritized for intensive follow-up observations, whereas unpromising interlopers that fail to meet the thresholds are abandoned. This approach minimizes the observational overhead by eliminating unnecessary follow-up for background events.

The performance of these color filters is quantified using two rigorous metrics, defined within the context of the observable sample set:
\begin{enumerate}
    \item \textbf{Recall:} The ratio of identified kilonova samples that pass the color cuts to the total number of identified kilonova samples in the simulation. This metric assesses the filter's completeness, specifically its ability to retain as many kilonova events as possible while discarding none.
    \item \textbf{Precision:} The ratio of identified kilonova samples that pass the color cuts to the total number of observable samples that satisfy the same criteria. This represents the purity of the resulting candidate list and evaluates the filter's power to reject afterglow-dominated false positives.
\end{enumerate}
By systematically varying the color filters and analyzing the resulting trade-off between Recall and Precision, we can determine the optimal selection criteria that maximize the scientific yield. Given the extreme rarity of these transients, our strategy prioritizes achieving a high sample completeness, enforcing a Recall exceeding the $90\%$ threshold while maximizing Precision to the greatest extent possible. These findings ultimately dictate the WFST multi-band observational strategy, ensuring that the telescope's cadence and filter choices are tuned to capture the red excess while efficiently managing limited operational resources.

\subsection{Band Selection and Survival Analysis} 
\begin{figure}[h!]
  \begin{subfigure}[h]{0.45\textwidth}
    \centering
    \includegraphics[width=\textwidth]{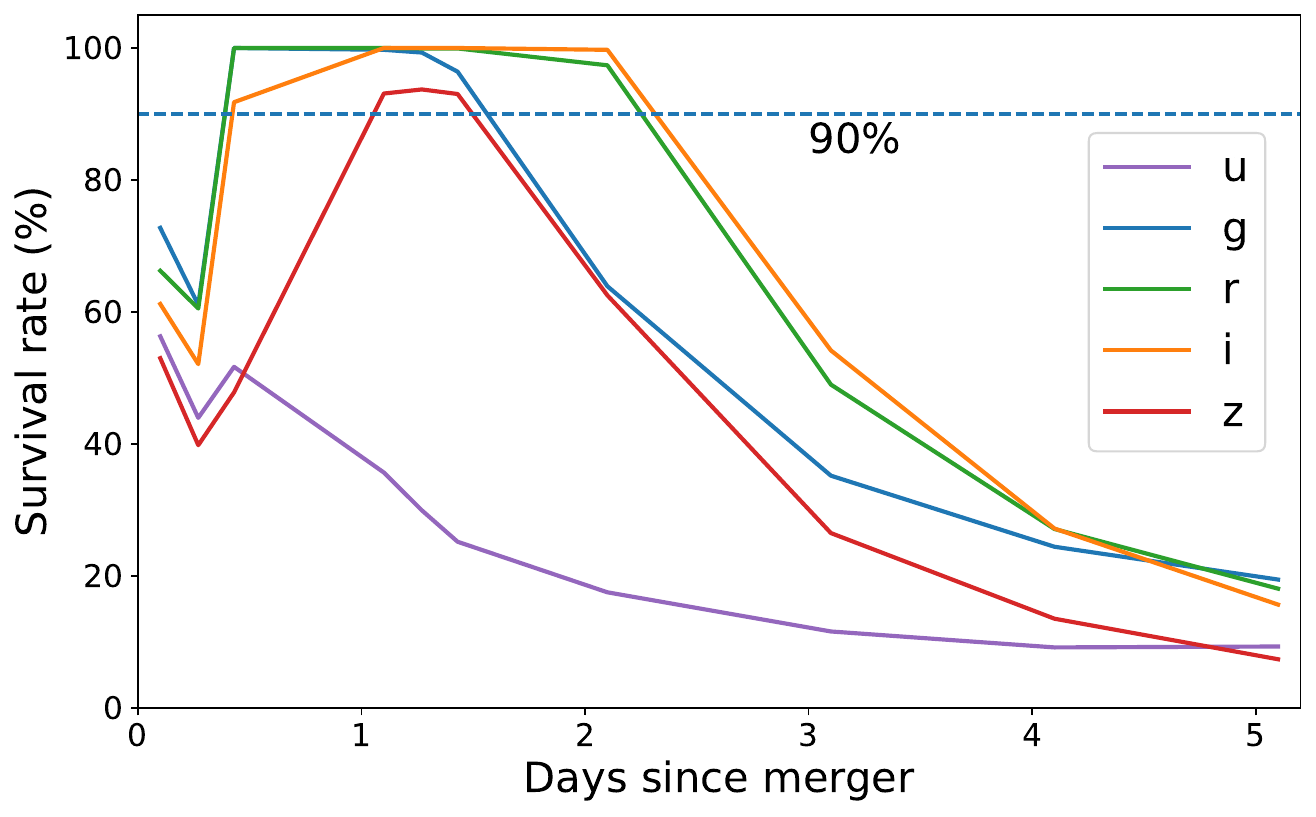}
    \caption{Single-band survival rate of Case 1. \label{fig:band170817}}
  \end{subfigure}  
  \hfill
  \begin{subfigure}[h]{0.45\textwidth}
    \centering
    \includegraphics[width=\textwidth]{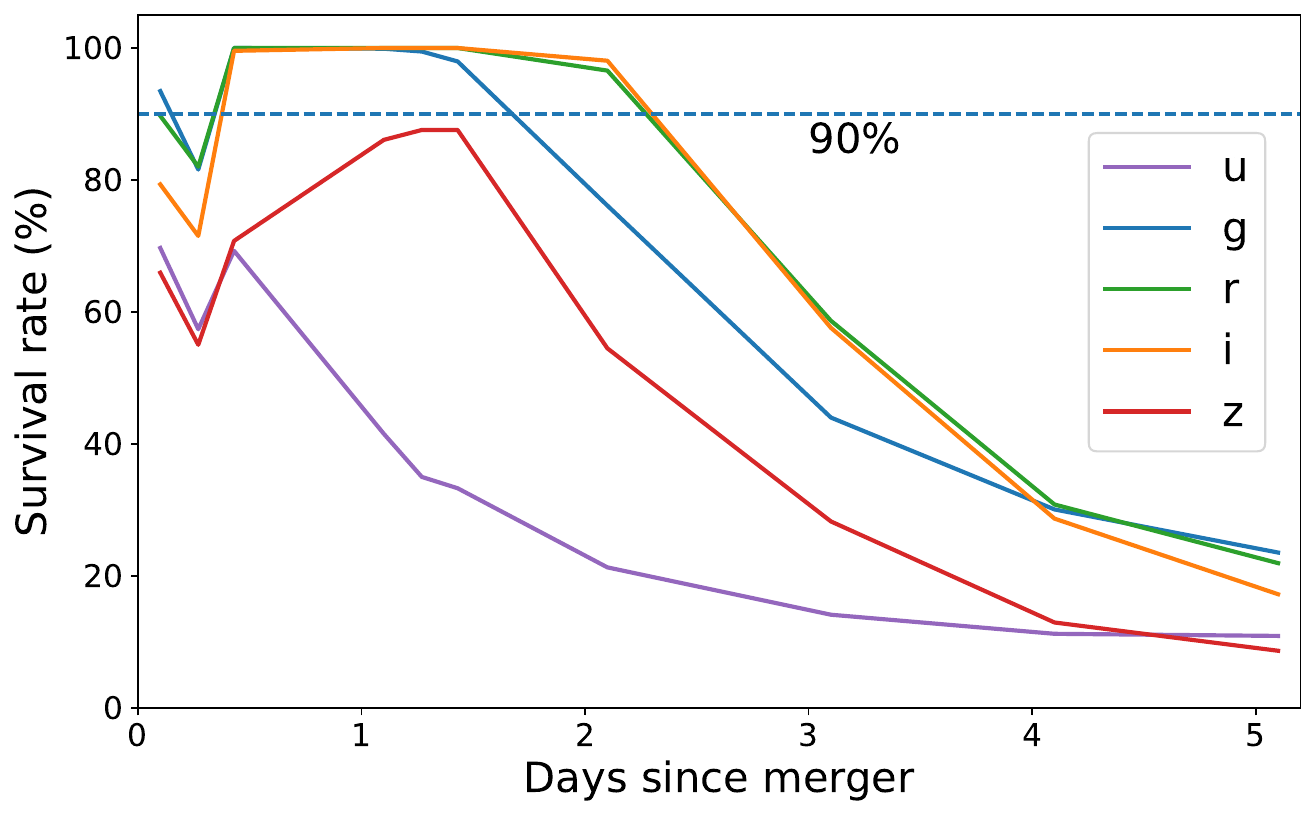}
    \caption{Single-band survival rate of Case 2. \label{fig:bandkilo}}
  \end{subfigure}
  \par\medskip
  \begin{subfigure}[h]{0.45\textwidth}
    \centering
    \includegraphics[width=\textwidth]{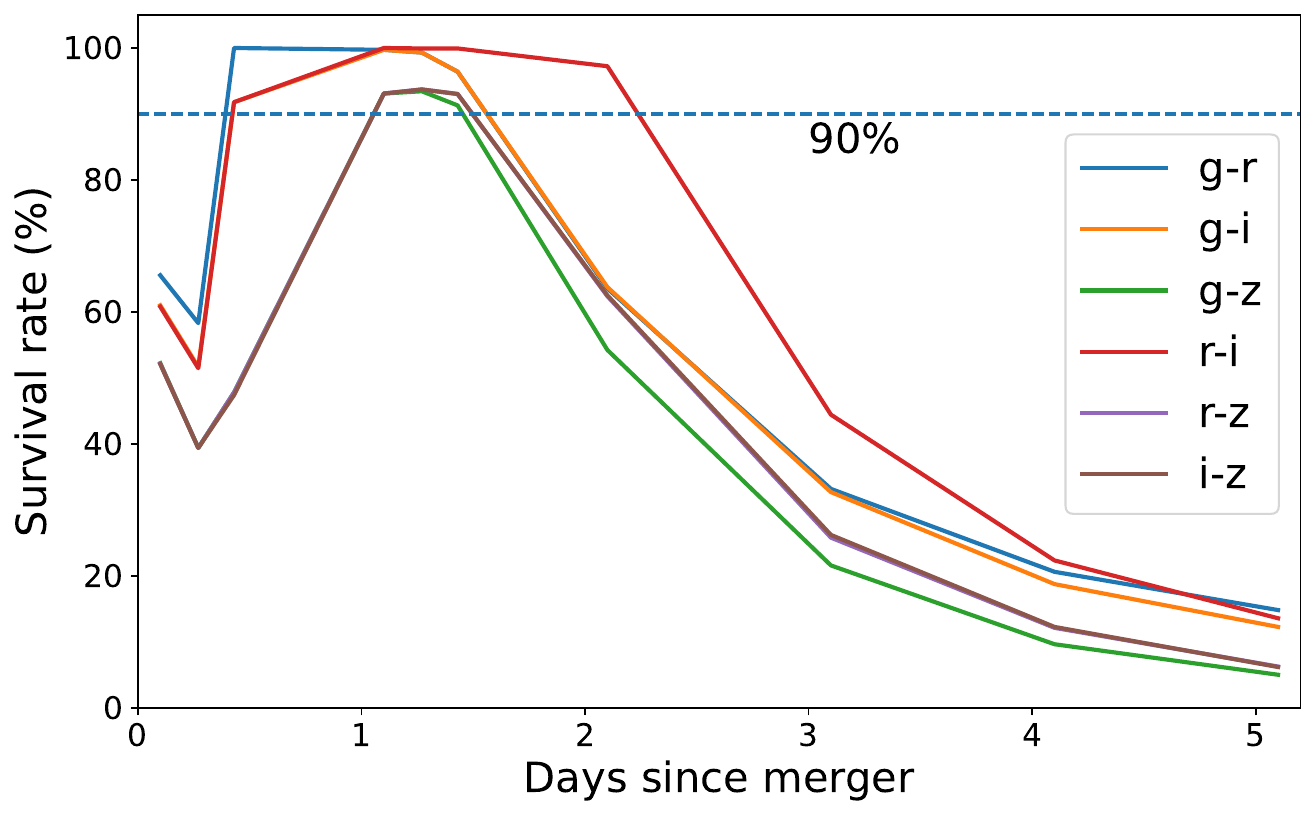}
    \caption{Dual-band survival rate of Case 1. \label{fig:bands170817}}
  \end{subfigure}
  \hfill
  \begin{subfigure}[h]{0.45\textwidth}
    \centering
    \includegraphics[width=\textwidth]{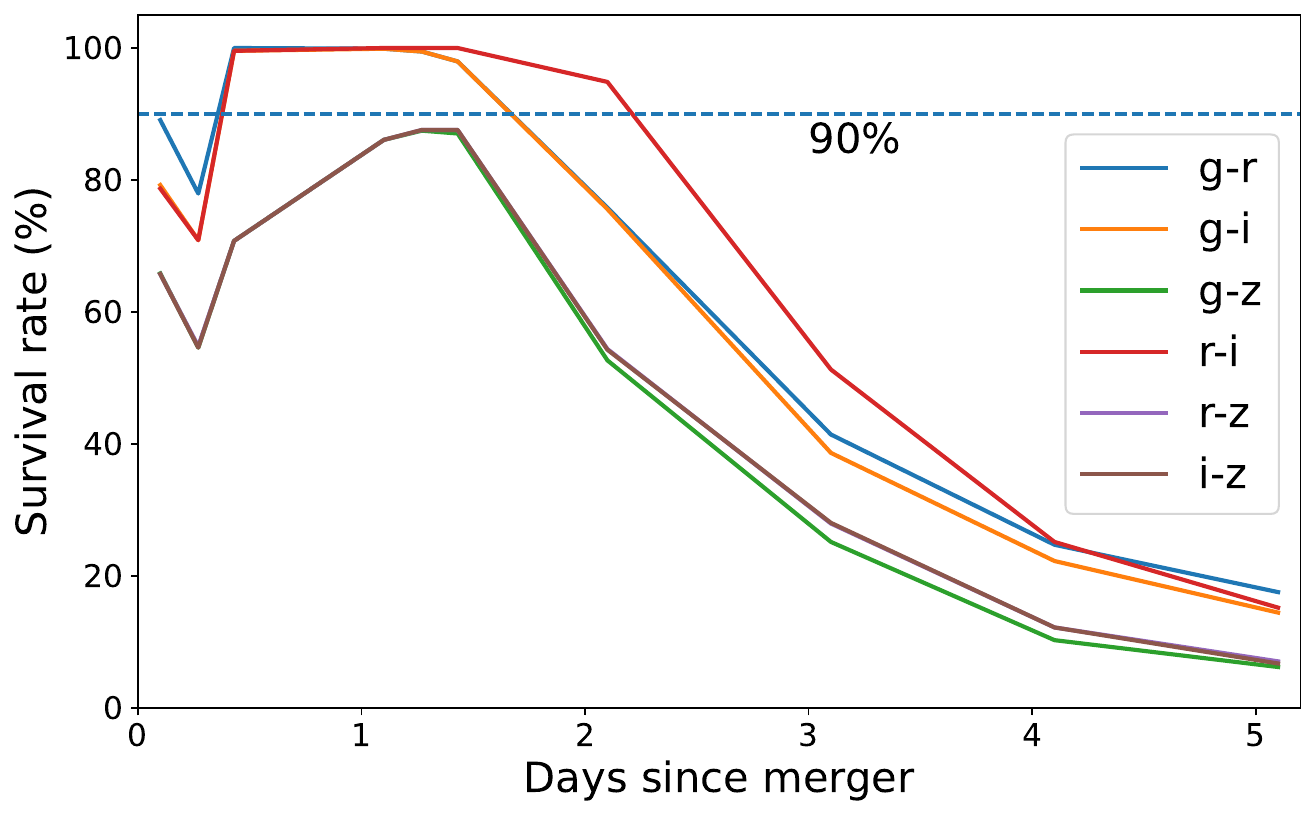}
    \caption{Dual-band survival rate of Case 2. \label{fig:bandskilo}}
  \end{subfigure}
  \caption{Survival rates of identified kilonova samples for two cases. Panels (a) and (b) present the single-band survival rates for Case 1 (template-based) and Case 2 (physically sampled), respectively. Panels (c) and (d) show the corresponding dual-band survival rates, essential for establishing color-selection criteria. The horizontal dashed line indicates the $90\%$ benchmark used for band selection. \label{fig:survival rate}}
\end{figure}

To optimize the color selection process, we first evaluate the band-specific survival rates of the identified kilonova samples, as shown in Figure~\ref{fig:survival rate}, which serves as a fundamental prerequisite for ensuring high recall. As illustrated in the single-band results (Figures~\ref{fig:band170817} and \ref{fig:bandkilo}), the $g$, $r$, and $i$ bands maintain consistently high survival rates, exceeding the $90\%$ threshold throughout the initial post-merger phase in both Case 1 and Case 2. Conversely, the $u$ band performs poorly in both scenarios, with survival rates falling below $60\%$ after the first night, leading us to exclude it from further color analysis. A notable difference emerges in the $z$ band: while it remains highly detectable in the template-based Case 1, its survival rate in the physically sampled Case 2 is initially suppressed during the first night, only stabilizing and approaching the $90\%$ benchmark after approximately one day as the ejecta expands and cools. The dual-band survival rates, which are critical for establishing reliable color indices, further dictate our observational cadence (Figures~\ref{fig:bands170817} and \ref{fig:bandskilo}). During the first night post-merger, only the $g-r$, $g-i$, and $r-i$ combinations consistently maintain survival rates above $90\%$ across both cases. Any pair involving the $z$ band (e.g., $g-z$, $r-z$) falls significantly short of this requirement during the first night, particularly in the more complex parameter space of Case 2. Consequently, our initial color screening for the first night relies exclusively on the $g$, $r$, and $i$ bands.

\subsection{Color Distributions and Component Separation}\label{subsec:color distributions}
\begin{figure}[h!]
  \centering
  \plottwo{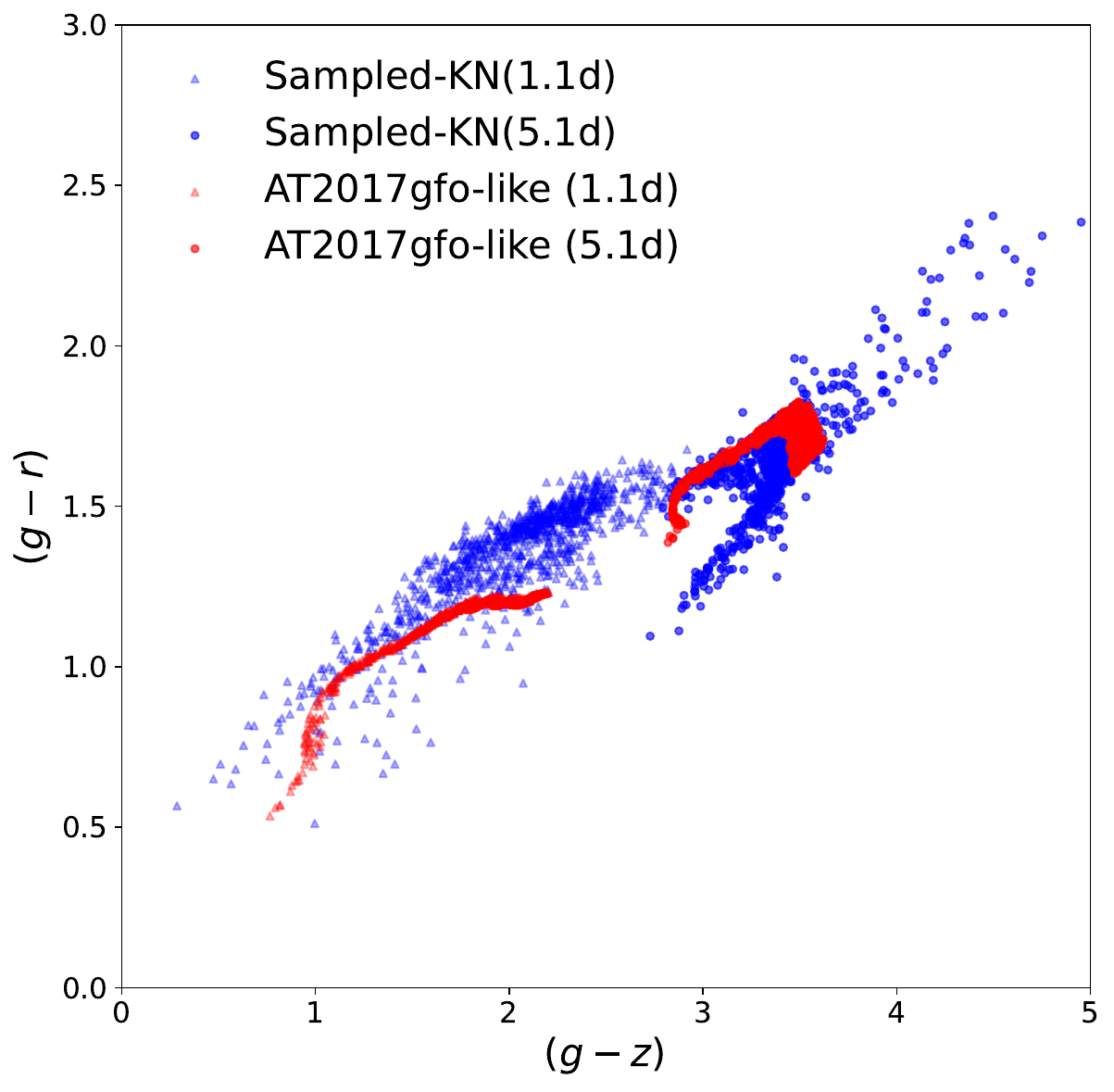}{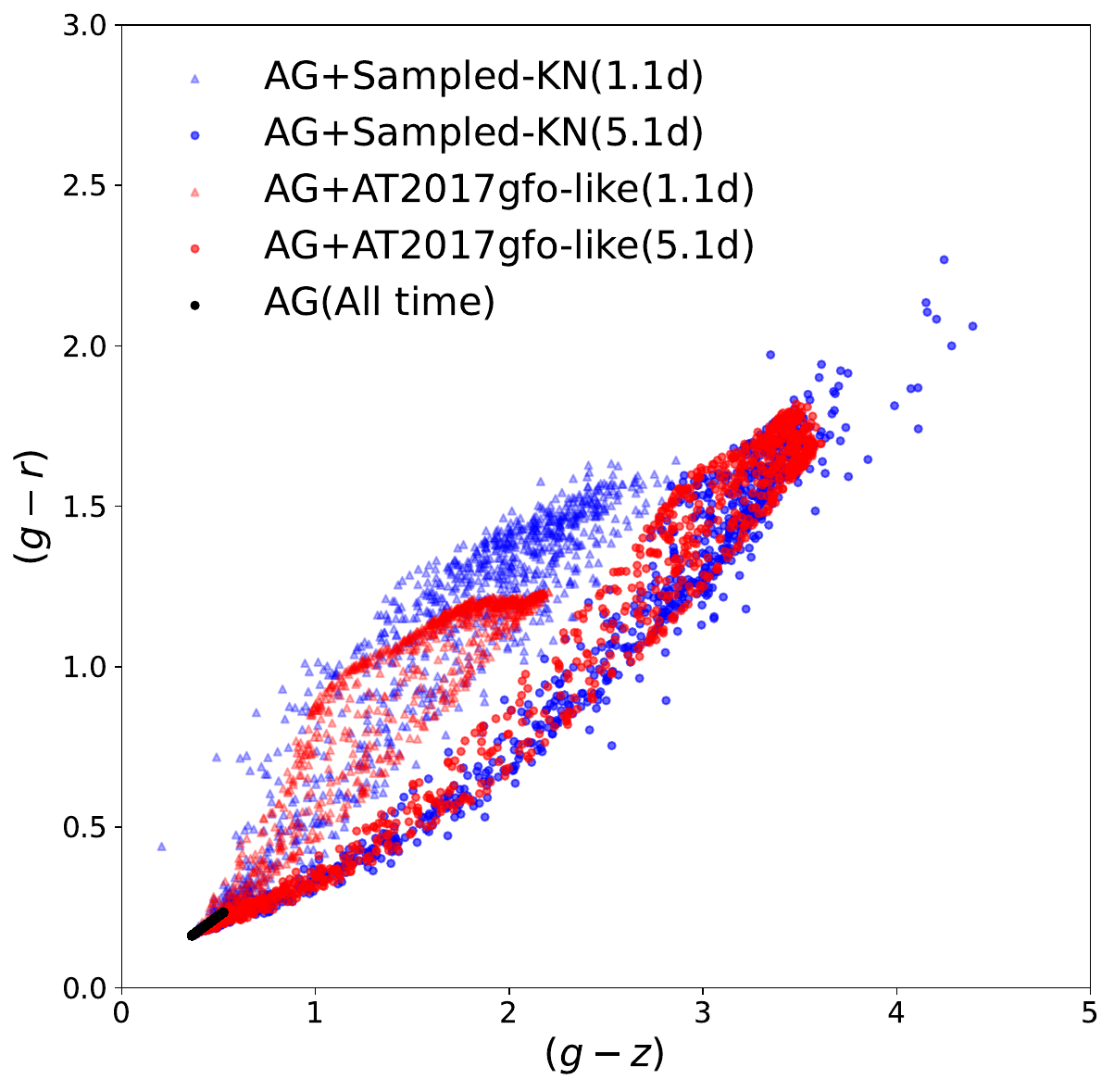}
  \caption{Color-color distributions in the $g-r$ versus $g-z$ plane for Night 2 (triangles) and Night 6 (dots). The plots compare the pure kilonova (left panel), and the pure afterglow and combined emission (right panel) for Case 1 (red) and Case 2 (blue). \label{fig:color}}
\end{figure}

As established, the initial color screening during the first night post-merger is restricted to the $g$, $r$, and $i$ bands due to survival rate constraints. From the second night onward, however, the $z$ band becomes highly operational. Therefore, we investigate the $z$-band color characteristics by studying the color-color distributions in the $g-r$ versus $g-z$ plane as a representative example. For visual clarity, we project a randomly selected subset of $1,000$ simulated samples onto this plane in Figure~\ref{fig:color}. We illustrate the color indices of the pure kilonova (\textit{left panel}) and the pure afterglow in comparison with the combined emission (\textit{right panel}) for both the template-based Case 1 (red) and the physically sampled Case 2 (blue) at $1.1\,\mathrm{d}$ and $5.1\,\mathrm{d}$ post-merger, using triangles and dots to denote the respective epochs.

The resulting distributions exhibit a distinct separation that is highly sensitive to the $r$-process elements in the kilonova ejecta. As shown in the left panel of Figure~\ref{fig:color}, the pure kilonovae consistently occupy a significantly red region; even at an early epoch of $1.1\,\mathrm{d}$, the samples already cluster at relatively red colors, and by $5.1\,\mathrm{d}$, the expansion and cooling of the ejecta drive the $g-z$ color index to values near $4.0$. The impact of model diversity is clearly visible in the broader dispersion of Case 2 compared to the tightly constrained tracks of Case 1, reflecting the wide range of physical parameters considered in our sampling. When the afterglow contamination is introduced (\textit{right panel}), the pure afterglow remains tightly clustered as a very blue population in the lower-left corner across all epochs (represented by the black points). At $1.1\,\mathrm{d}$ (triangles), the combined emission already deviates from this pure afterglow cluster owing to the presence of the kilonova component. At $5.1\,\mathrm{d}$ (dots), despite the presence of the afterglow background which pulls the total observed colors slightly bluer, the kilonova-dominated samples still maintain a distinct separation from the pure afterglow in the color-color space.

Consequently, these findings demonstrate that the inclusion of the $z$ band provides an ideal diagnostic environment with a clear, color-dependent separation boundary. While a bright afterglow can contaminate the signal, the substantial red excess captured by the $z$ band remains highly resilient against background interlopers. We therefore conclude that incorporating the $z$ band into the WFST observational framework, starting from the second night, is uniquely suited to capture this pronounced red excess, which serves as a definitive signature for efficient kilonova identification.

\subsection{Dual-Band and Triple-Band Selection Efficiency}
\begin{figure}[h!]
  \begin{subfigure}[h]{0.44\textwidth}
    \centering
    \includegraphics[width=\textwidth]{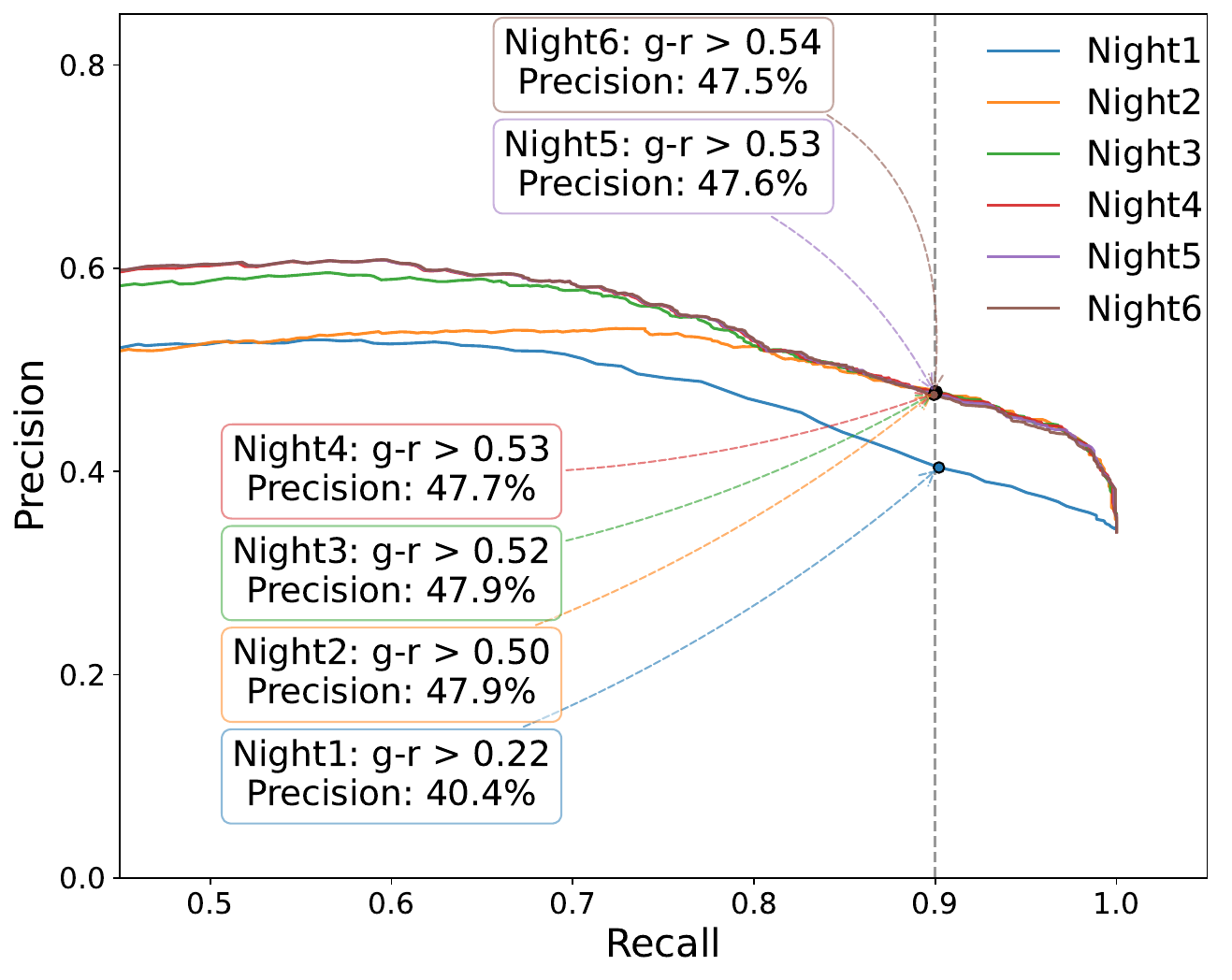}
    \vspace{-15pt}
    \caption{$g-r$ (Case 1). \label{fig:gr170817}}
  \end{subfigure}
  \hspace{1cm}  
  \begin{subfigure}[h]{0.44\textwidth}
    \centering
    \includegraphics[width=\textwidth]{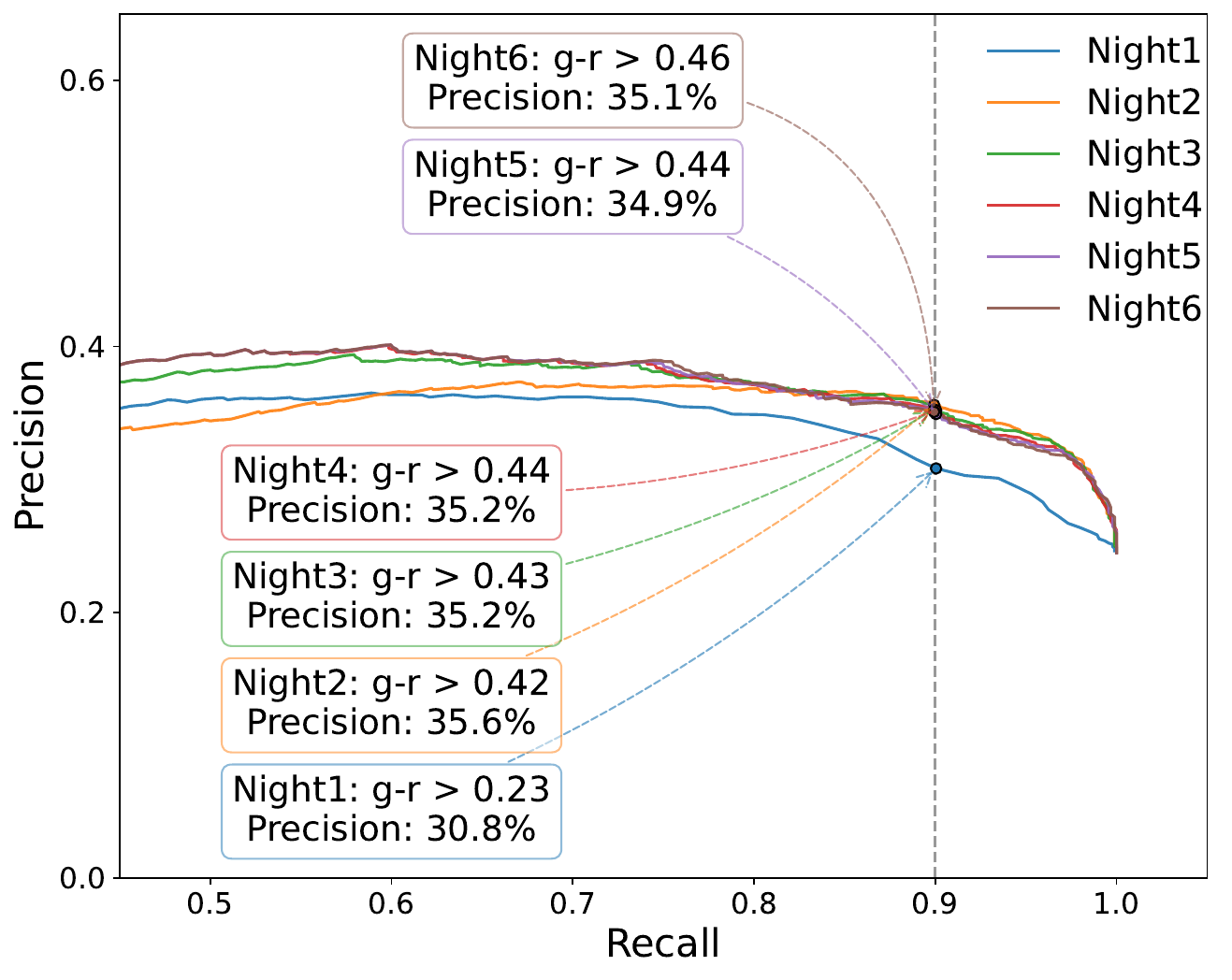}
    \vspace{-15pt}
    \caption{$g-r$ (Case 2). \label{fig:grkn}}
  \end{subfigure} 
  \par\smallskip
  \begin{subfigure}[h]{0.44\textwidth}        
    \centering
    \includegraphics[width=\textwidth]{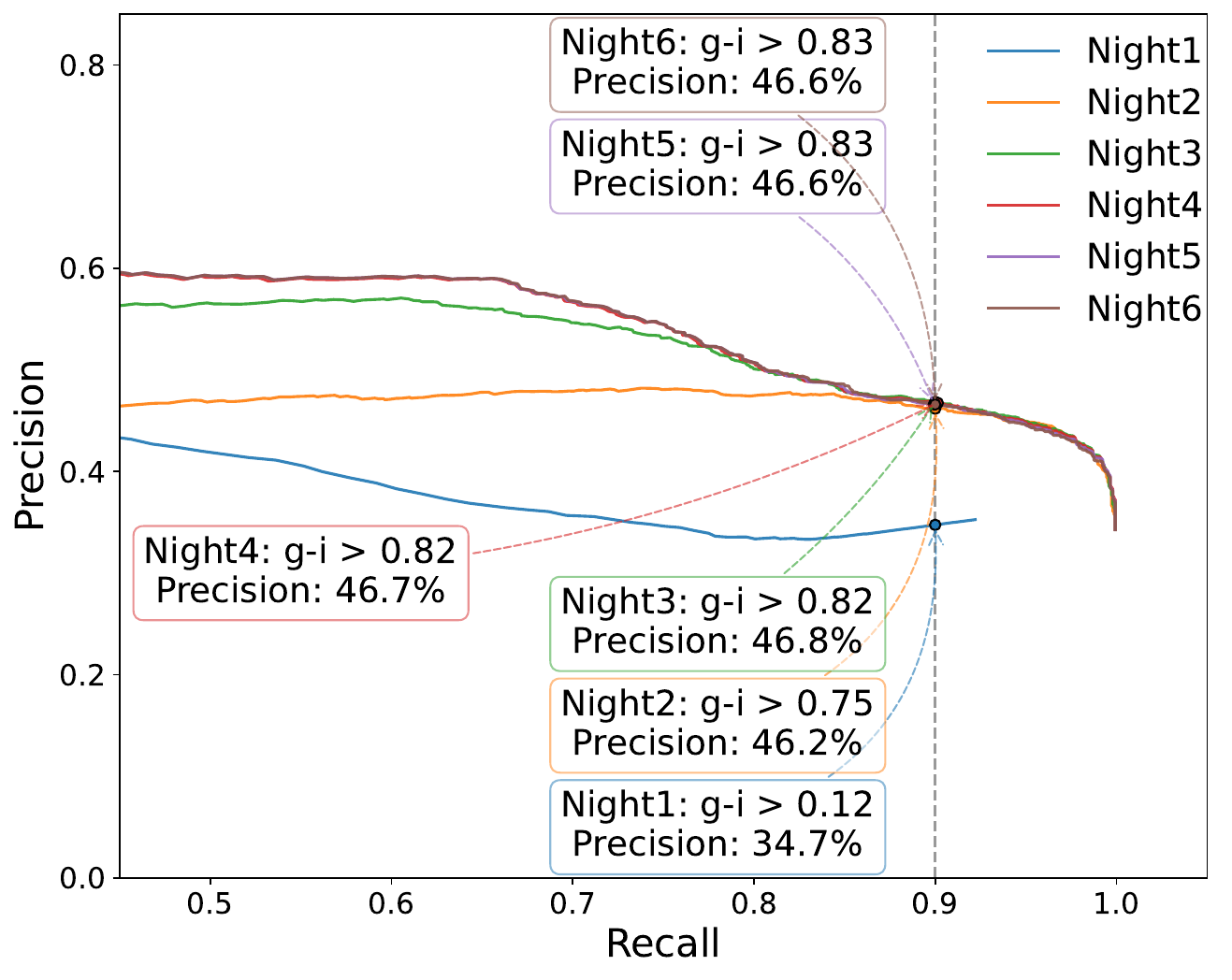}
    \vspace{-15pt}
    \caption{$g-i$ (Case 1). \label{fig:gi170817}}
  \end{subfigure}
  \hspace{1cm}  
  \begin{subfigure}[h]{0.44\textwidth}
    \centering
    \includegraphics[width=\textwidth]{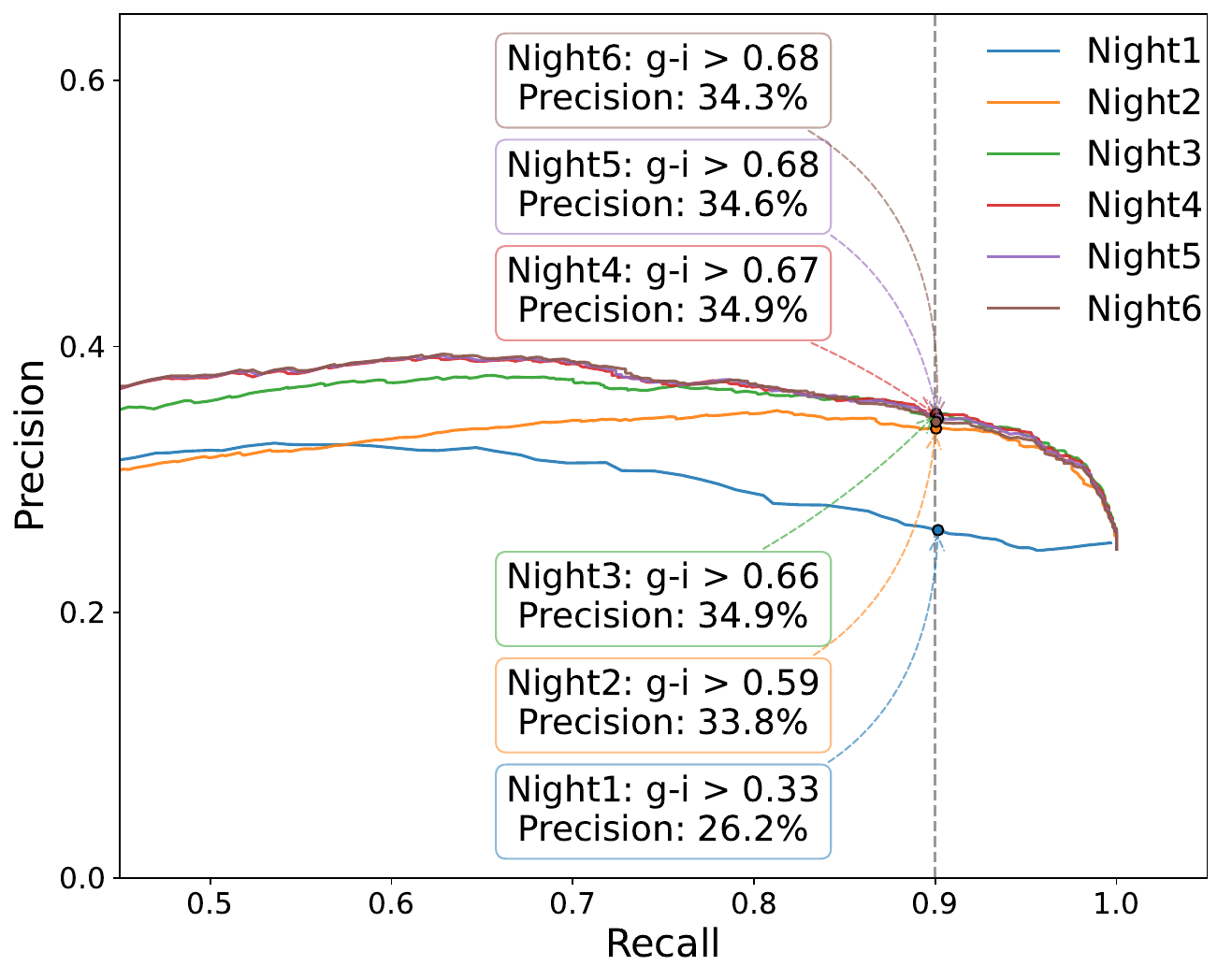}
    \vspace{-15pt}
    \caption{$g-i$ (Case 2). \label{fig:gikn}}
  \end{subfigure} 
  \par\smallskip
  \begin{subfigure}[h]{0.44\textwidth}
    \centering
    \includegraphics[width=\textwidth]{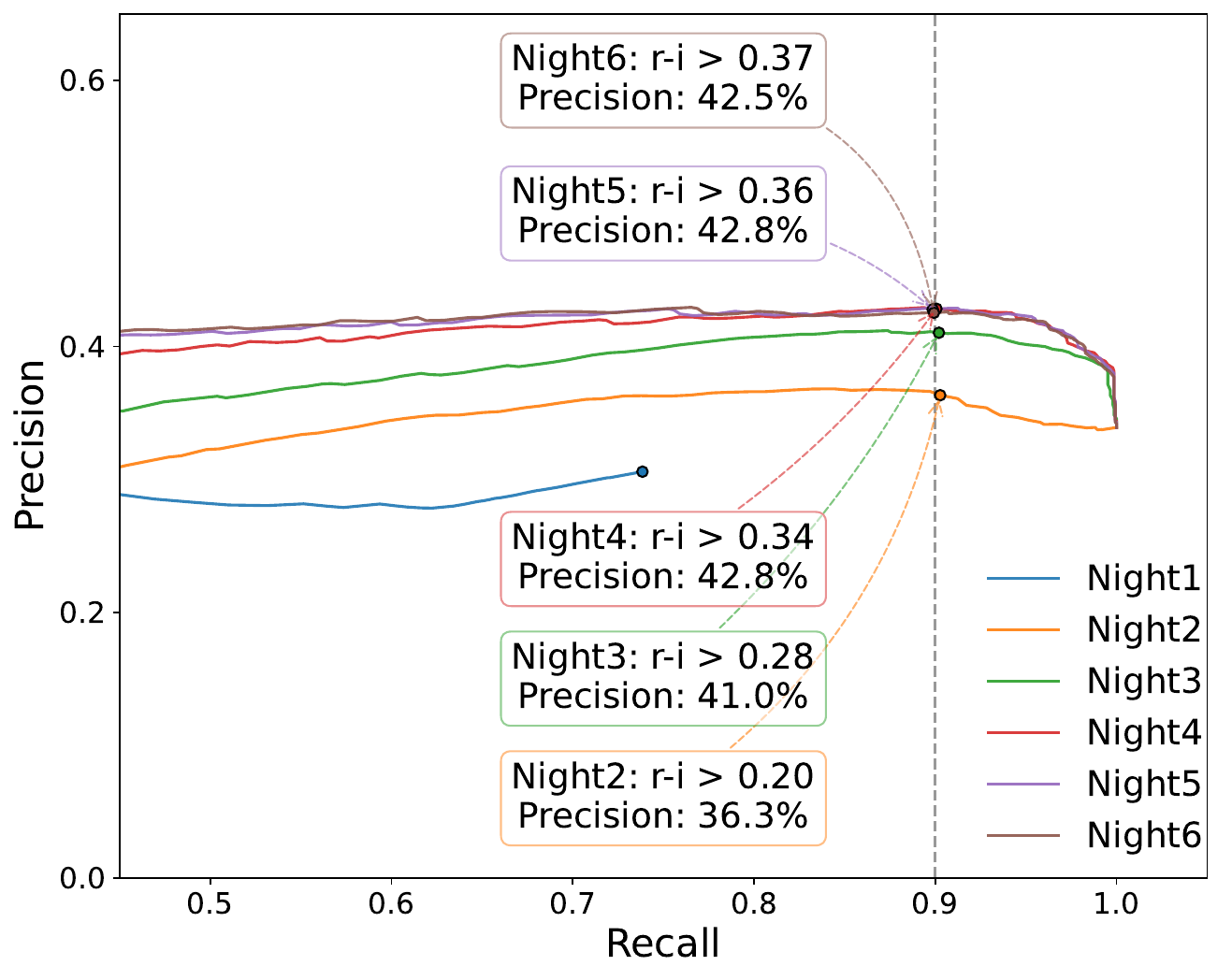}
    \vspace{-15pt}
    \caption{$r-i$ (Case 1). \label{fig:ri170817}}
  \end{subfigure}
  \hspace{1cm}  
  \begin{subfigure}[h]{0.44\textwidth}  
    \centering
    \includegraphics[width=\textwidth]{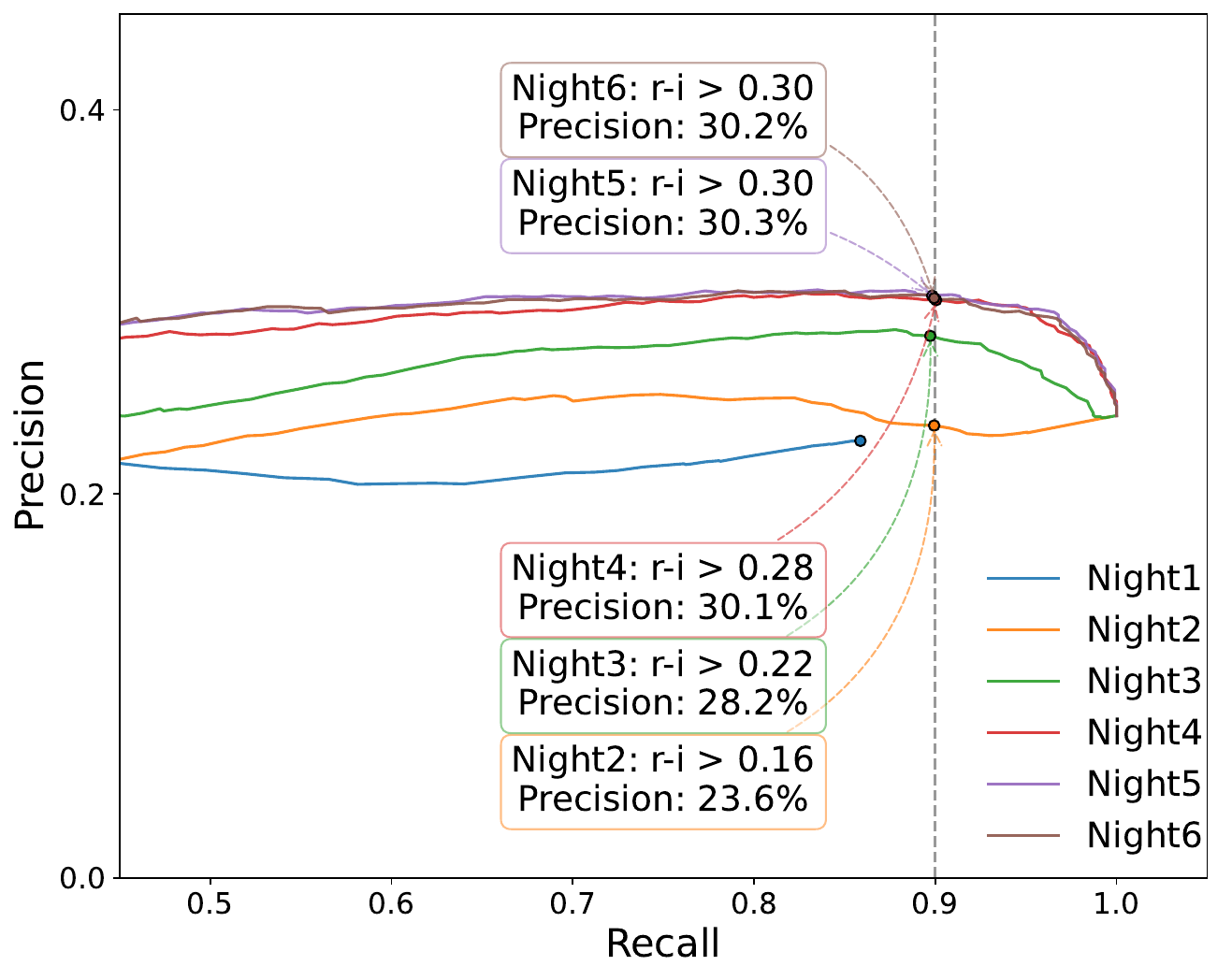}
    \vspace{-15pt}
    \caption{$r-i$ (Case 2). \label{fig:rikn}}
  \end{subfigure}
  \vspace{-5pt}
  \caption{Precision-Recall (PR) analysis for cumulative dual-band color selection across Case 1 (left column) and Case 2 (right column). Panels (a) and (b) compare the $g-r$ combination; panels (c) and (d) compare $g-i$; and panels (e) and (f) compare $r-i$. Each panel illustrates the performance evolution from Night 1 to Night 6, with colors representing different observational epochs. The annotated boxes indicate the Precision and the specific color threshold required to maintain a benchmark Recall of $0.9$ (indicated by the vertical black dashed line). \label{fig:2bands color}}
\end{figure}

Our analysis is performed on a cumulative basis, where the data for any given Night n incorporates all observational information from Night 1 through Night n. To establish a robust identification criterion, we utilize the maximum color observed for each transient as the primary screening parameter. Subsequently, the resulting Recall and Precision are evaluated for each individual color cut. The performance of different dual-band combinations for the template-based (Case 1) and physically sampled (Case 2) populations is quantified using Precision-Recall (PR) curves, with the results for $g-r$, $g-i$, and $r-i$ illustrated in Figure~\ref{fig:2bands color}. Across both scenarios, the $g-r$ combination emerges as the most potent diagnostic tool for early-time kilonova identification (Figures~\ref{fig:gr170817} and \ref{fig:grkn}). In Case 1, $g-r$ achieves a Precision of $40.4\%$ on Night 1, rapidly saturating by Night 2 at $47.9\%$ (with a cut of $g-r>0.50$). A similar evolutionary trend is observed in Case 2, where Precision improves from $30.8\%$ on Night 1 to a stable $35.6\%$ by Night 2 (with a cut of $g-r>0.42$). While the $g-i$ index (Figures~\ref{fig:gi170817} and \ref{fig:gikn}) demonstrates a comparable saturation pattern, its peak Precision remains consistently lower than that of $g-r$. The $r-i$ index exhibits the poorest performance across all tests (Figures~\ref{fig:ri170817} and \ref{fig:rikn}), which is particularly evident on Night 1, as it fails to reach the $0.9$ Recall threshold and is thus excluded from the point-wise comparison. This diminished efficiency stems from the similar magnitude evolution and timescales of kilonovae in the $r$ and $i$ bands, resulting in insufficient color contrast to effectively separate them from afterglow backgrounds.

The transition from the AT2017gfo template (Case 1) to a physically sampled population (Case 2) results in a general decrease in peak precision, reflecting the increased complexity and broader parameter space. Despite this reduction in absolute purity, the robustness of the $g-r$ signature remains a universal feature. This confirms that even under varied physical conditions—such as different ejecta masses—the early-time $g-r$ color provides a definitive and reliable indicator of kilonova emergence. The consistent results from both cases reinforce a prioritized observational strategy for the WFST. During the critical $32-48\,\mathrm{h}$ post-merger (Night 1 and Night 2), resources should be concentrated on high-cadence monitoring in the $g$ and $r$ bands.

\begin{figure}[h!]
  \begin{subfigure}[h]{0.46\textwidth}
    \centering
    \includegraphics[width=\textwidth]{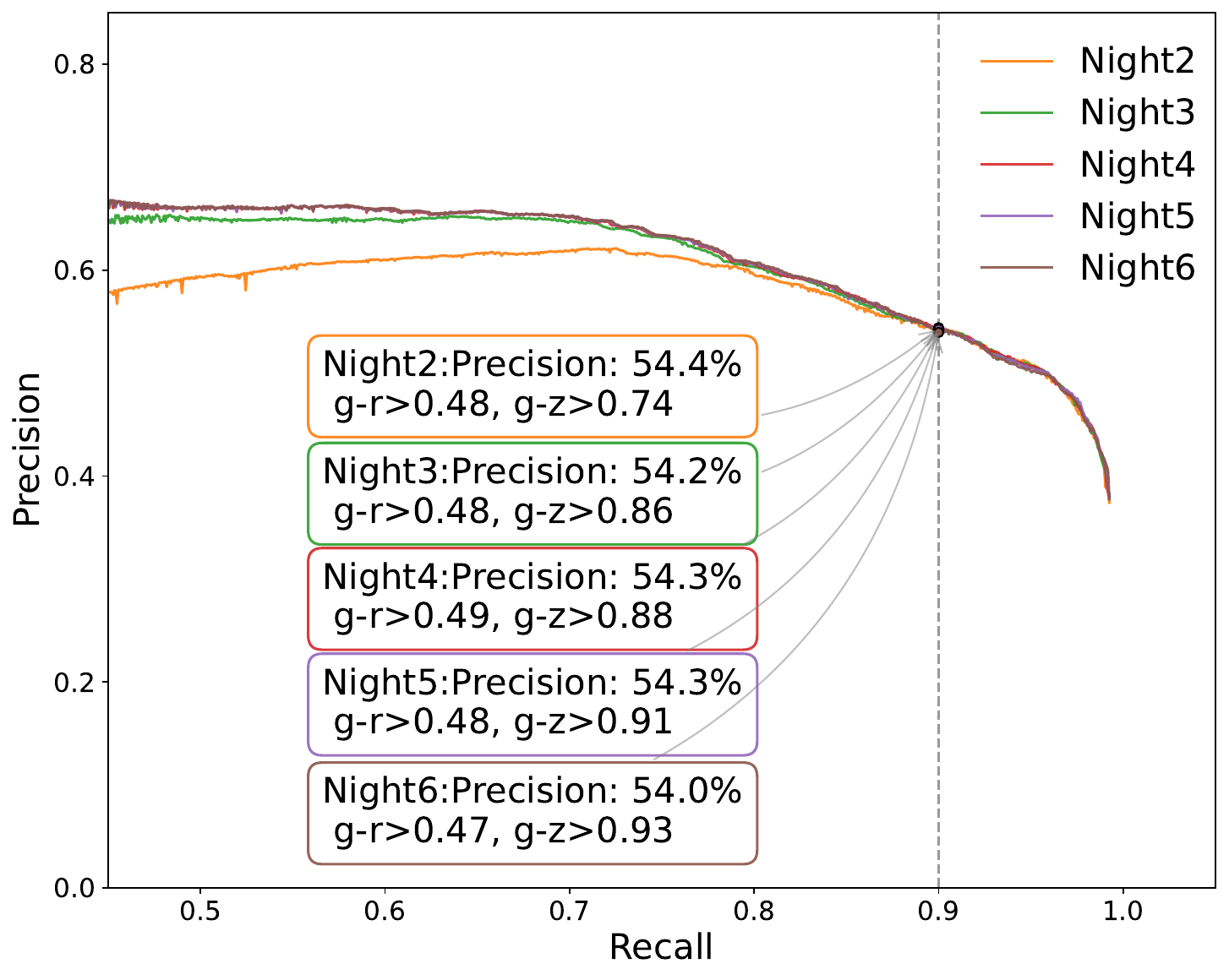}
    \caption{$g-r \& g-z$ (Case 1). \label{fig:grz170817}}
  \end{subfigure}
  \hspace{1cm} 
  \begin{subfigure}[h]{0.46\textwidth}
    \centering
    \includegraphics[width=\textwidth]{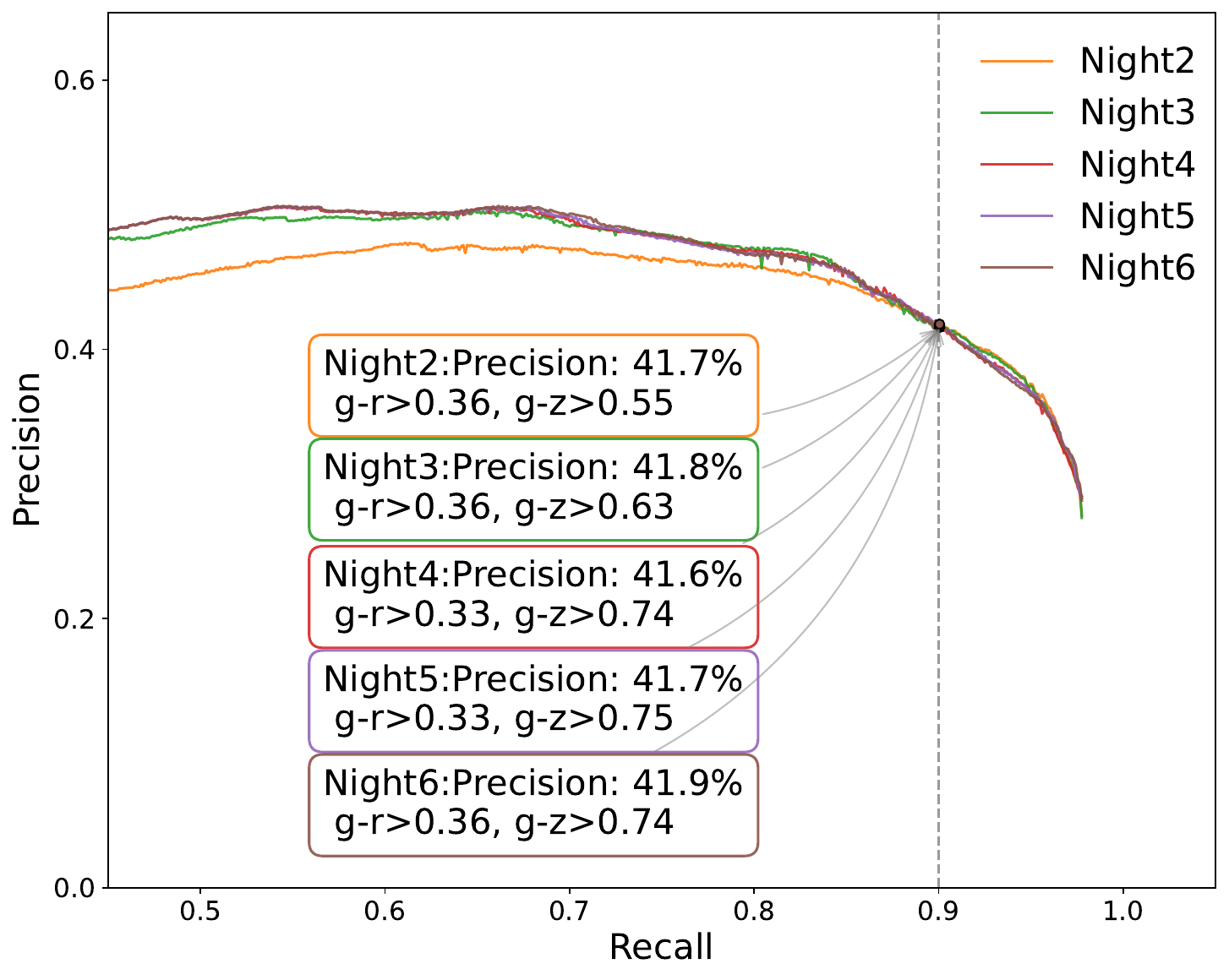}
    \caption{$g-r \& g-z$ (Case 2). \label{fig:grzkn}}
  \end{subfigure} 
  \caption{Precision-Recall (PR) analysis for cumulative triple-band ($g, r, z$) joint selection in Case 1 (a) and Case 2 (b). The analysis begins from Night 2, following the integration of $z$-band data. Annotated boxes highlight the Precision and the dual-threshold color cuts ($g-r$ and $g-z$) required to maintain a benchmark Recall of $0.9$ (indicated by the vertical black dashed line). \label{fig:triple_band_pr}}
\end{figure}

To overcome the Precision plateau inherent in dual-band filtering and further elevate sample purity, we investigate a triple-band joint selection strategy by integrating the $z$ band starting from the second night post-merger. Driven by the color separation demonstrated in the $g-z$ versus $g-r$ plane in Section~\ref{subsec:color distributions}, this sequential approach leverages the $z$-band availability to apply secondary, tighter color constraints. The cumulative PR curves for both cases are illustrated in Figure~\ref{fig:triple_band_pr}.

The inclusion of the $z$ band significantly elevates the Precision of the cumulative identification process compared to the dual-band results across both scenarios. For Case 1 (Figure~\ref{fig:grz170817}), upon the introduction of the $z$ band on Night 2, the Precision at a benchmark Recall of $0.9$ jumps to $54.4\%$. This represents a substantial improvement over the $47.9\%$ maximum precision achieved by the $g-r$ dual-band strategy alone. For Case 2 (Figure~\ref{fig:grzkn}), the joint criteria yield a Precision of $41.7\%$ on Night 2, marking a notable increase from the $35.6\%$ Precision ceiling observed in the preceding dual-band analysis. In both cases, the triple-band performance demonstrates rapid convergence and remains exceptionally stable from Night 2 through Night 6, with Precision fluctuating minimally ($54\%$--$55\%$ for Case 1 and $41\%$--$42\%$ for Case 2). This high stability indicates that the critical discriminatory color information is effectively captured as soon as the $z$ band is incorporated. To maintain a consistently high Recall, the optimal color cuts are tailored to each specific epoch. In Case 2, the selection thresholds shift from ($g-r > 0.36$, $g-z > 0.55$) on Night 2 to more stringent criteria of ($g-r > 0.36$, $g-z > 0.74$) by Night 6. This adjusted $g-z$ threshold quantitatively aligns with the intrinsically redder color characteristics exhibited by the kilonova ejecta at later stages. While the Precision in Case 2 is lower than that in Case 1 due to the expanded parameter space, the relative efficiency gain provided by the triple-band strategy remains remarkably consistent.

\subsection{Robustness to Initial Observation Time Delay}
We further test the robustness of our color-based identification method against delays in the initial observation time, using simulations with initial epochs $t_0=0.2~\mathrm{d}$, $0.5~\mathrm{d}$, and $0.8~\mathrm{d}$. The Precision at a fixed recall of $0.9$ benchmark for all color combinations and all epochs (Case 1) is summarized in Table \ref{tab:precision}. We find that the relative performance ranking of the color indices is identical across all initial observation times: the triple-band $g-r-z$ combination consistently achieves the highest Precision, followed by $g-r$, $g-i$, and $r-i$. Notably, our $t_0=0.1~\mathrm{d}$ setup provides a conservative estimate of the identification performance compared to slightly delayed observations. This is because kilonova ejecta are still relatively blue at very early times, and the characteristic red color signature strengthens over the first $0.5~\mathrm{d}$ as the ejecta expand and cool, improving the color contrast with the bluer afterglow component. The evolution of Precision across nights differs between early and delayed setups, which is a natural consequence of the rapid early reddening of kilonova ejecta. In the first ~0.5 days after merger, the expanding material cools quickly and its optical color shifts rapidly from blue to red. For the earliest initial epoch $0.1~\mathrm{d}$ and $0.2~\mathrm{d}$, the kilonova is still relatively blue for the first night, resulting in weaker color contrast against the afterglow background and lower selection precision. Precision then rises substantially by Night 2 as the characteristic red signature develops. For setups starting at $0.5\ \mathrm{d}$ or later, the first observational night already captures the epoch where the $g-r$ color contrast is near its maximum, so precision is close to its peak already on Night 1 and changes only modestly thereafter. For all tested initial epochs, the Precision of $g-r \& g-z$ converges rapidly and stabilizes by Night 2, with only minimal fluctuations thereafter.\textcolor{red}{
\begin{table}[htbp]
\centering
\renewcommand{\arraystretch}{0.9}
\caption{Precision for $0.9$ Recall benchmark from PR analysis of cumulative color selections (Case 1) of four initial epochs $t_0=0.1~\mathrm{d}$, $0.2~\mathrm{d}$, $0.5~\mathrm{d}$, $0.8~\mathrm{d}$ for Night1-Night6. \label{tab:precision}}
\vspace{6pt}
\begin{tabular}{c | c c c c c c c} 
$t_0$ & colors & Night1 & Night2 & Night3 & Night4 & Night5 & Night6 \\
\hline
\multirow{4}{*}{0.1d} 
& $g-r$ & 40.4\% & 47.9\% & 47.9\% & 47.7\% & 47.6\% & 47.5\% \\
& $g-i$ & 34.7\% & 46.2\% & 46.8\% & 46.7\% & 46.6\% & 46.6\% \\
& $r-i$ & \nodata & 36.3\% & 41.0\% & 42.8\% & 42.8\% & 42.5\% \\
& $g-r \& g-z$ & \nodata & 54.4\% & 54.2\% & 54.3\% & 54.3\% & 54.0\% \\
\cline{1-8}
\multirow{4}{*}{0.2d} 
& $g-r$ & 46.1\% & 51.0\% & 51.1\% & 51.0\% & 50.8\% & 50.6\%\\
& $g-i$ & 35.1\% & 48.9\% & 49.4\% & 49.4\% & 49.4\% & 49.5\% \\
& $r-i$ & \nodata& 38.0\% & 42.4\% & 43.9\% & 44.0\% & 43.7\% \\
& $g-r \& g-z$ & \nodata& 58.5\% & 58.9\% & 58.8\%& 58.8\% & 58.7\%\\
\cline{1-8}
\multirow{4}{*}{0.5d} 
& $g-r$ & 48.9\% & 49.5\% & 49.1\%& 48.8\% & 48.8\% &48.8\% \\
& $g-i$ & 42.3\% & 47.1\% & 46.8\%& 46.7\% & 46.8\% & 47.1\%\\
& $r-i$ & \nodata& 36.9\% & 39.6\%& 40.5\% & 40.2\% & 40.0\%\\
& $g-r \& g-z$ & \nodata& 56.0\% & 56.0\%& 55.8\% & 56.0\% & 56.3\%\\
\cline{1-8}
\multirow{4}{*}{0.8d} 
& $g-r$ & 49.3\% & 47.1\% & 47.8\% & 47.5\% & 47.0\% & 47.1\%\\
& $g-i$ & 43.8\% & 46.0\% & 45.5\% & 45.3\% & 45.2\% & 45.2\%\\
& $r-i$ & \nodata& 36.9\% & 38.8\% & 39.2\% & 39.1\% & 39.0\%\\
& $g-r \& g-z$ & \nodata& 51.9\% & 52.0\% & 51.6\% & 51.3\% & 52.0\%\\
\end{tabular}
\end{table}
}

These results validate our proposed observational strategy for the WFST, emphasizing a prioritized band allocation that aligns with the intrinsic color characteristics of kilonova signals. Given the rapid convergence of identification efficiency observed across both cases and the full range of practical initial observation delays, the discriminative power of the filters stabilizes as early as the second night post-merger. We therefore propose a specific operational cadence for multi-band surveys under resource-limited conditions: resources should be concentrated on high-cadence $g$ and $r$ band monitoring during the first night to exploit the immediate color signatures and guarantee a high Recall, followed by the mandatory integration of the $z$ band from the second night onward. This transition from dual-band ($g-r$) to triple-band ($g-r$ \& $g-z$) joint analysis effectively breaks the Precision limitations inherent in early observations. These filters rule out afterglow-dominated false positives with maximum efficiency, ensuring that spectroscopic resources and multi-messenger characterizations are reserved exclusively for high-potential kilonova candidates.

\section{Conclusions and Outlook} \label{sec:conclusion}
In this study, we have developed a comprehensive numerical framework to evaluate the identification potential of the WFST within a sGRB-triggered ToO observational strategy, focusing on kilonovae embedded within afterglow emission. By simulating $10,000$ multi-messenger transients for both template-based and physically sampled BNS merger scenarios, we have quantified the observational constraints and established an optimized identification pipeline.

Our analysis reveals that kilonova identification with the WFST is primarily limited by distance rather than afterglow microphysics. While the diverse population in Case 2 exhibits a lower absolute identification efficiency due to intrinsically fainter candidates, it shows a physical response to parameter variations that is remarkably consistent with Case 1. In both scenarios, the identification efficiency remains robust, staying above $80\%$ within approximately $600\,\mathrm{Mpc}$ for AT2017gfo-like events. This demonstrates a robust resilience to variations in jet energy, circumburst density, or other intrinsic afterglow parameters. This confirms that the distinct features of a kilonova can be reliably disentangled from a wide range of afterglow environments using our Fisher Information Matrix framework. Furthermore, we find that the transition from a specific template to a randomized physical population results in an approximately $35\%$ reduction in the expected annual identification count. This discrepancy highlights the critical necessity of accounting for intrinsic kilonova diversity when projecting survey yields. 
Based on current volumetric BNS merger rates and within the Fermi-GBM sGRB-triggered ToO observing scenario, we estimate that the WFST is capable of identifying approximately $0.3^{+0.5}_{-0.2}$ to $0.4^{+0.8}_{-0.3}$ kilonovae per year within a $1,400\,\mathrm{Mpc}$ volume in the optimistic scenario. Nevertheless, these numbers stem from simplified analytical estimations and do not fully capture the complexity of realistic survey operations.

Importantly, through a rigorous Precision-Recall analysis, we demonstrate that the discriminative power of color-based filters exhibits a rapid saturation effect. Both dual-band ($g-r$) and triple-band ($g-r\,\&\,g-z$) identification efficiencies reach a stable plateau as early as the second night post-merger. This early saturation indicates that the most critical chromatic information for breaking the afterglow-kilonova degeneracy is captured within the first $48~mathrm{h}$. Consequently, we propose a staged observation strategy: prioritizing high-cadence $g$- and $r$-band monitoring during the first night, followed by the mandatory integration of the $z$ band from the second night. Through multi-band joint analysis for the first two nights, candidates that fail to exhibit definitive multi-band color signatures are flagged as unpromising interlopers, allowing the automated scheduling system to terminate their subsequent tracking. This approach effectively optimizes the limited observational resources of the telescope while maintaining high completeness. It is worth noting that while our color-based identification efficiency reaches saturation by the second night, the three-night detection requirement remains a necessary conservative constraint. This ensures that candidates identified via color filters are not only physically consistent with kilonova models but also possess sufficient temporal coverage for robust light-curve characterization.  We acknowledge that our simulation of an initial observation time at $t=0.1 \mathrm{d}$ represents an optimistic scenario. The potential loss of very early data could affect the completeness of the kilonova light curves and consequently impact their robust identification. To verify the impact of initial observation timing, we perform additional robustness simulations with three delayed initial epochs: $0.2~\mathrm{d}$, $0.5~\mathrm{d}$, and $0.8~\mathrm{d}$. The number of reliably identified kilonovae decreases moderately with increasing delay, which is physically expected. But our identification strategy relies dominantly on post-peak photometric features, so a moderate delay does not substantially degrade the discriminative power. As for the color identification framework, the relative superiority of the triple-band $g-r-z$ strategy as well as the staged observational strategy remain valid across all tested initial epochs.

Looking forward, the synergy between optical surveys and multi-wavelength observations will be crucial for further refining kilonova identification. While this work focuses on the optical regime, the integration of high-energy gamma-ray triggers, X-ray counterparts (to constrain afterglow energetics), and late-time radio observations (to probe the total kinetic energy of the jet) will provide a more holistic view of the merger event. Indeed, as a preliminary step toward this panchromatic synergy, we investigate the joint observation of optical and X-ray bands in Appendix~A, demonstrating that such a combination can enhance kilonova identification. We also extend our analysis in Appendix~B to incorporate magnetar-powered kilonovae—a distinct subclass driven by the rotational energy loss of a millisecond magnetar—which further demonstrates the robustness and adaptability of our conclusions to diverse central engine physics. Future refinements to this framework could incorporate these multi-band datasets into a joint Bayesian or machine learning architecture to achieve even higher purity in candidate catalogs. Ultimately, the WFST is poised to be a cornerstone instrument in the multi-messenger era, significantly enhancing our ability to characterize the electromagnetic counterparts of BNS mergers and explore the synthesis of heavy elements in the universe.
\begin{acknowledgments}
We would like to thank Cong Zhou, Huiyu Wang, Ken Chen, Lei He and Rui Niu for useful suggestions. This work is supported by the National Natural Science Foundation of China (grant Nos. 12325301 and 12273035), Strategic Priority Research Program of the Chinese Academy of Science (grant No. XDB0550300), the National Key R\&D Program of China (grant Nos. 2021YFC2203102 and 2022YFC2204602). 
\end{acknowledgments}

\facilities{WFST, SWIFT/XRT}

\software{
     AFTERGLOWPY \citep{ryan2020gamma}, 
     ASTROPY \citep{robitaille2013astropy, astropy2018astropy, astropy2022astropy}, 
     MATPLOTLIB \citep{hunter2007matplotlib}, 
     NUMPY \citep{harris2020array}, 
     PANDAS \citep{mckinney2010proceedings, reback2020pandas}, 
     SCIPY \citep{virtanen2020scipy}, 
     SNCOSMO \citep{barbary2016sncosmo},
     REDBACK \citep{sarin2024redback}.
}
\bibliography{main}{}
\bibliographystyle{aasjournalv7}

\appendix\label{sec:appendix}
\section{Contribution of X-rays to Kilonova Identification}\label{xray} 
\setcounter{table}{0}
\renewcommand{\thetable}{A\arabic{table}}
\setcounter{figure}{0}
\renewcommand{\thefigure}{A\arabic{figure}}
Beyond optical light curves, X-ray emissions serve as a critical diagnostic tool for kilonova identification \citep{tanvir2013kilonova, troja2019afterglow, jin2020kilonova, yang2022long, yang2024lanthanide}, as it provides independent and robust constraints on afterglow physics. Since kilonovae are predominantly thermal transients radiating in the optical and infrared regimes, the observed X-ray flux is produced exclusively by the non-thermal afterglow component. Incorporating X-ray data allows for the effective decoupling of the afterglow background from the total multi-wavelength flux, thereby significantly enhancing the precision of kilonova parameter extraction.

\begin{figure}[h!]
  \begin{subfigure}[h]{0.45\textwidth}
    \centering
    \includegraphics[width=\textwidth]{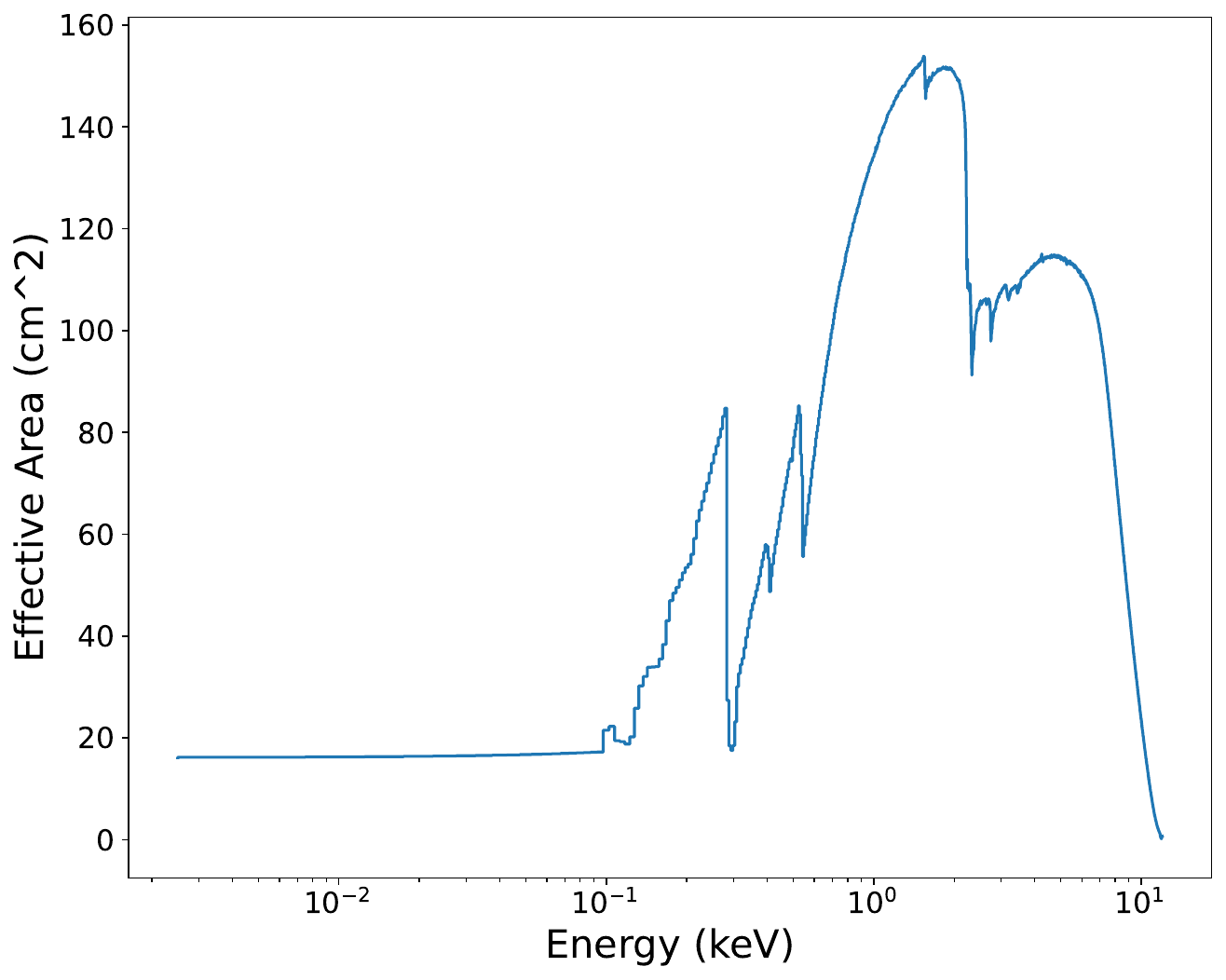}
    \caption*{}
  \end{subfigure}
  \hfill  
  \begin{subfigure}[h]{0.45\textwidth}
    \centering
    \includegraphics[width=\textwidth]{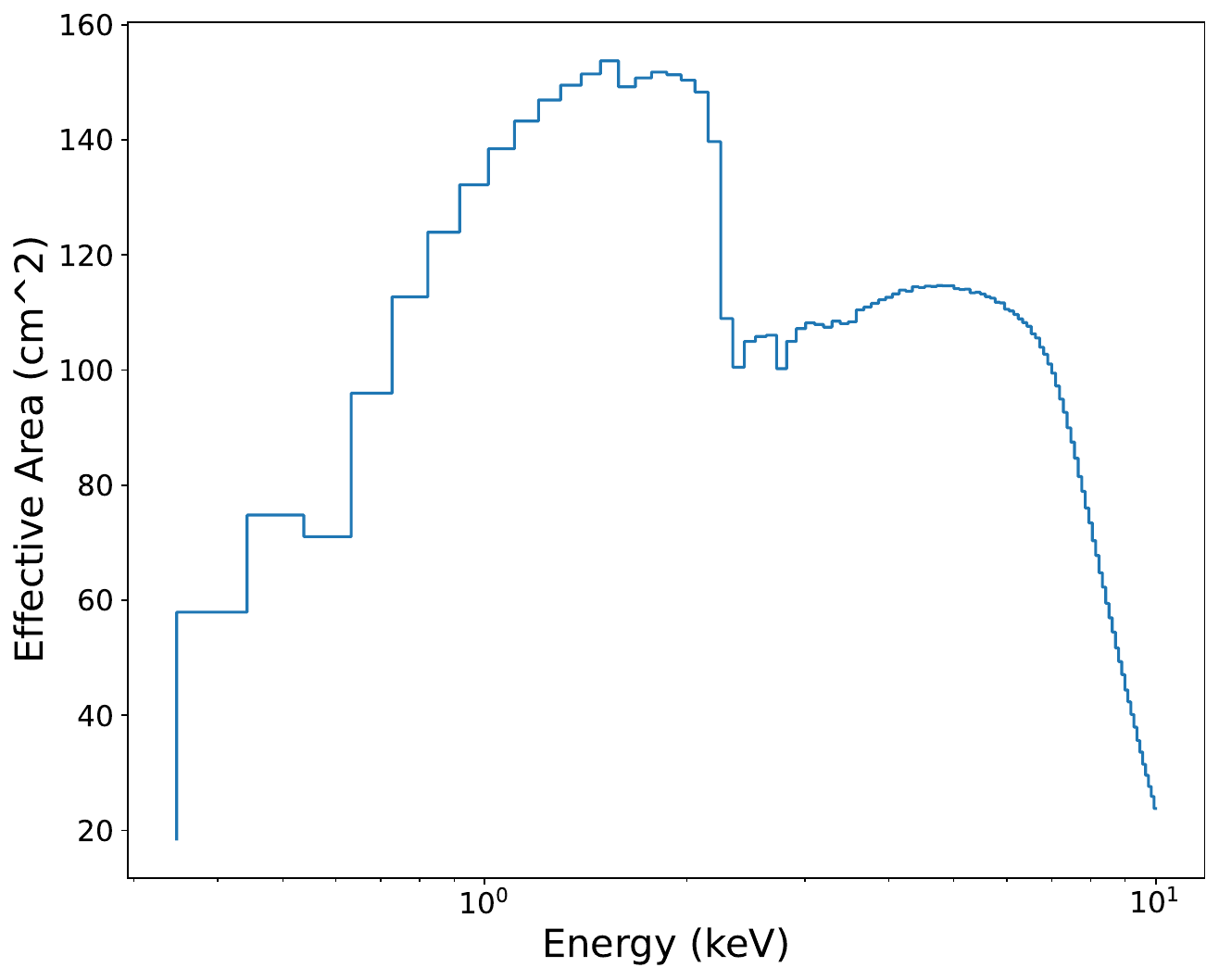}
    \caption*{}
  \end{subfigure} 
  \caption{Swift/XRT ARF visualization; left: original ARF, right: simplified ARF. \label{fig:arf}}
\end{figure}

We simulate X-ray observations based on the performance of the Swift X-ray Telescope (XRT) \citep{burrows2005swift} operating in Photon Counting (PC) mode, following the standardized protocols of the UK Swift Science Data Centre (UKSSDC) \citep{evans2007online, evans2009methods}. The Ancillary Response File (ARF) utilized for our simulations is illustrated in Figure~\ref{fig:arf}. To optimize computational efficiency while maintaining spectral integrity, we utilize a simplified ARF (consisting of $103$ discrete energy points) to normalize the X-ray flux within the $0.3$--$10\,\mathrm{keV}$ range. The simulated observation window extends from $100\,\mathrm{s}$ to $1\,\mathrm{d}$ post-merger, with $10$ epochs sampled log-uniformly. The resulting observational data, including photon counts and flux uncertainties, are generated assuming a Poisson distribution. We apply a flux sensitivity threshold of $10^{-14}\, \mathrm{erg}\, \mathrm{s}^{-1}\, \mathrm{cm}^{-2}$, consistent with the typical deep-limit capability of the Swift/XRT.

\begin{figure}[h!]
  \centering
  \includegraphics[width=1\textwidth]{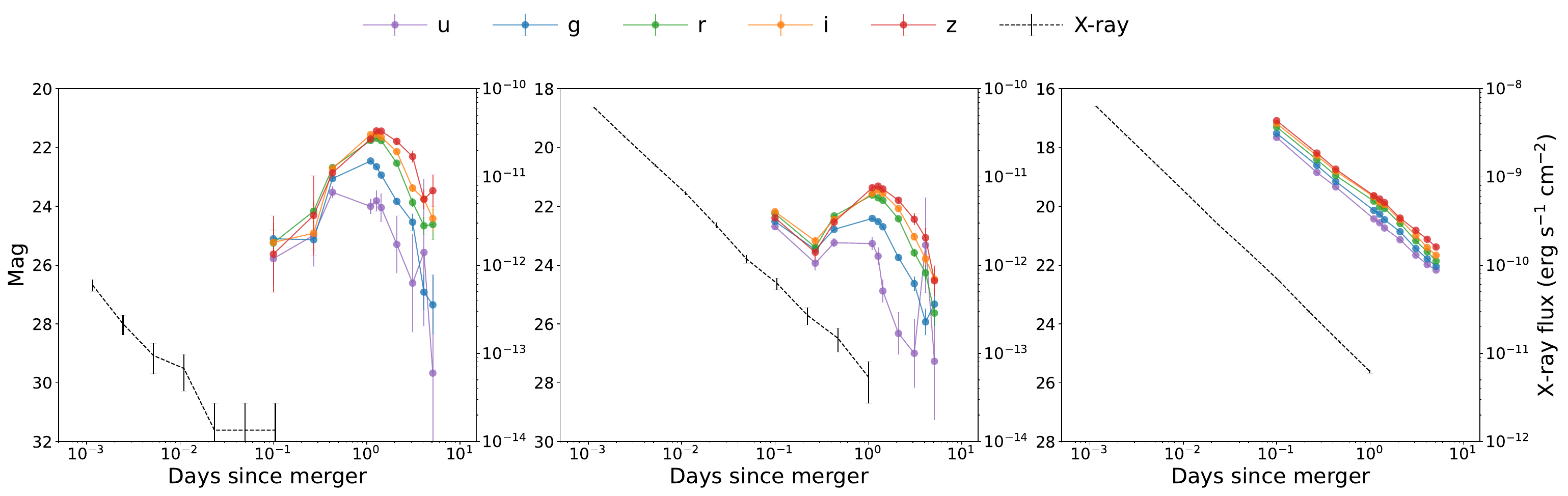}
  \caption{Multi-band light curves of three representative simulated samples. The solid lines with error bars denote the total observed magnitude in each band ($u, g, r, i, z$), while the black dashed lines represent the X-ray flux. All samples are placed at a fixed distance of $400\,\mathrm{Mpc}$. The kilonova parameters are fixed to the best-fit values of the AT2017gfo template: $m_{\mathrm{ej}}^{\mathrm{dyn}} = 10^{-2.27} M_{\odot}$, $m_{\mathrm{ej}}^{\mathrm{wind}} = 10^{-1.28} M_{\odot}$, $\Phi = 49.5\,^\circ$ \citep{dietrich2020multimessenger}. The afterglow parameters are set to the median values of our adopted distributions: $\theta_{\mathrm{obs}}=21.09\,^\circ$, $\log n_0=-2.5$, $p=2.35$, $\theta_c=0.25$, $\log \epsilon_e=-2.0$, $\log \epsilon_B=-3.5$, while $\log E_0$ takes values of $49.0, 50.5, 52.0$ from left to right, respectively. \label{fig:xray samples}} 
\end{figure}

Owing to the statistical independence of X-ray and optical observations, the total FIM is calculated as the sum of their individual contributions:
\begin{equation}
F_{ij, \mathrm{total}} = F_{ij, \mathrm{optical}} + F_{ij, \mathrm{X-ray}}.
\end{equation}
This joint analysis markedly suppresses parameter uncertainties. As demonstrated in Figure~\ref{fig:xray samples}, the inclusion of X-ray constraints reduces the uncertainty of a representative sample from $\sigma_A=0.111$ to $0.081$ in the left panel and from $\sigma_A=0.088$ to $0.082$ in the middle panel. In the right panel, where the afterglow is inherently dominant, the uncertainty $\sigma_A=0.222$ remains largely unchanged, illustrating the physical limits of identification.

Overall, the projected identification efficiency for the WFST is significantly improved by the addition of X-ray data. For Case 1, the number of identified samples increases from 1,449 to 1,854, raising the projected identified count from $N_{\mathrm{id}}=0.4^{+0.8}_{-0.3}$ to $0.6^{+1.0}_{-0.4}$ per year. Similarly, for the physically sampled Case 2, identified samples increase from 934 to 1,482, elevating the annual identified count from $N_{\mathrm{id}}=0.3^{+0.5}_{-0.2}$ to $0.4^{+0.8}_{-0.3}$. These results underscore the importance of rapid X-ray follow-up in maximizing the scientific yield of BNS merger surveys.

\section{Magnetar-powered Kilonovae Identification}\label{magnetar} 
\setcounter{table}{0}
\renewcommand{\thetable}{B\arabic{table}}
\setcounter{figure}{0}
\renewcommand{\thefigure}{B\arabic{figure}}
Besides standard radioactive-powered kilonovae, a distinct subclass termed magnetar-powered kilonova emerges when the merger remnant evolves into a long-lived supramassive NS with ultra-strong dipolar magnetic fields (millisecond magnetar) \citep{yu2013bright, Sarin_2022}. These transients exhibit optical luminosities significantly higher than those of radioactive-powered kilonovae, occasionally reaching magnitudes comparable to supernovae. Unlike radioactive transients, where the luminosity drops rapidly as the energy release from $r$-process isotopes fades, magnetar-powered kilonovae are characterized by sustained energy injection. Specifically, while the radioactive heating rate diminishes sharply as the ejecta expands and the radioactive material decays—leading to a steep decline in the light curve—the central magnetar engine acts as a continuous power source. This long-term energy contribution effectively counteracts the intrinsic decay, resulting in a significantly flatter light curve during the late stages of the transient. 

In the specific case of AT2017gfo, a hybrid model, which combined r-process radioactive heating with a magnetar central engine, was employed to characterize its multi-wavelength light curves \citep{yu2018long}. A pivotal component of this framework is the integration of a magnetar, whose rotational energy loss is primarily driven by GW braking. The study demonstrates that superimposing the magnetar spin-down luminosity upon the standard radioactive heating effectively mitigates the limitations of purely radioactive frameworks in explaining the prolonged light curve. 

The physical evolution of the kilonova in this model is defined by several key parameters. The initial energy injection is defined by the magnetar's dipole luminosity $\xi L_{\mathrm{md}}(0)$, while its temporal duration and the decay of the light curve are dictated by the spin-down timescale $t_{\mathrm{sd}}$. The total ejecta mass $M_{\mathrm{ej}}$ and the ejecta opacity $\kappa$ regulate the photometric evolution, with the latter primarily determined by the abundance of heavy r-process elements. Furthermore, the dynamical expansion of the material is defined by the velocity range between $v_{\mathrm{min}}$ and $v_{\mathrm{max}}$, while the density profile index $\delta$ characterizes the radial mass distribution and governs the migration of the photosphere through the ejecta. Collectively, these parameters define the physical state of the post-merger system and dictate the resulting multi-wavelength photometric signature.   
\begin{deluxetable*}{llc}
\tablewidth{0pt}
\tablecaption{Parameters of the magnetar-powered kilonova model from \citet{yu2018long}. \label{tab:kn parameters}}
\tablehead{\colhead{Parameter} & \colhead{Meaning} & \colhead{Distribution}}
\startdata
$\xi L_{\rm md}(0)$ & Initial energy injection ($10^{41}\,\mathrm{erg}\ \mathrm{s}^{-1}$) & $U(0.1,\,100)$ \\
$t_{\rm sd}$ & Spin-down time scale ($10^{5}\mathrm{s}$) & $\mathcal{U}(0.01,\,1)$ \\
$M_{\rm ej}$ & Total ejecta mass ($0.01\,M_{\odot}$) & $\mathcal{U}(0.1,\,10)$ \\
$\kappa$ & Radiative opacity ($\mathrm{g}\ \mathrm{cm}^{-2}$) & $\mathcal{U}(0.1,\,10)$ \\
$v_{\rm min}$ & Minimum expansion velocity ($c$)& $\mathcal{U}(0.01,\,0.15)$ \\
$v_{\rm max}$ & Maximum expansion velocity ($c$)& $\mathcal{U}(0.18,\,0.40)$ \\
$\delta$ & Ejecta density profile index & $\mathcal{U}(1.0,\,3.0)$ \\
\enddata
\end{deluxetable*}

\begin{figure*}[h!]
\centering
\includegraphics[width=1\textwidth]{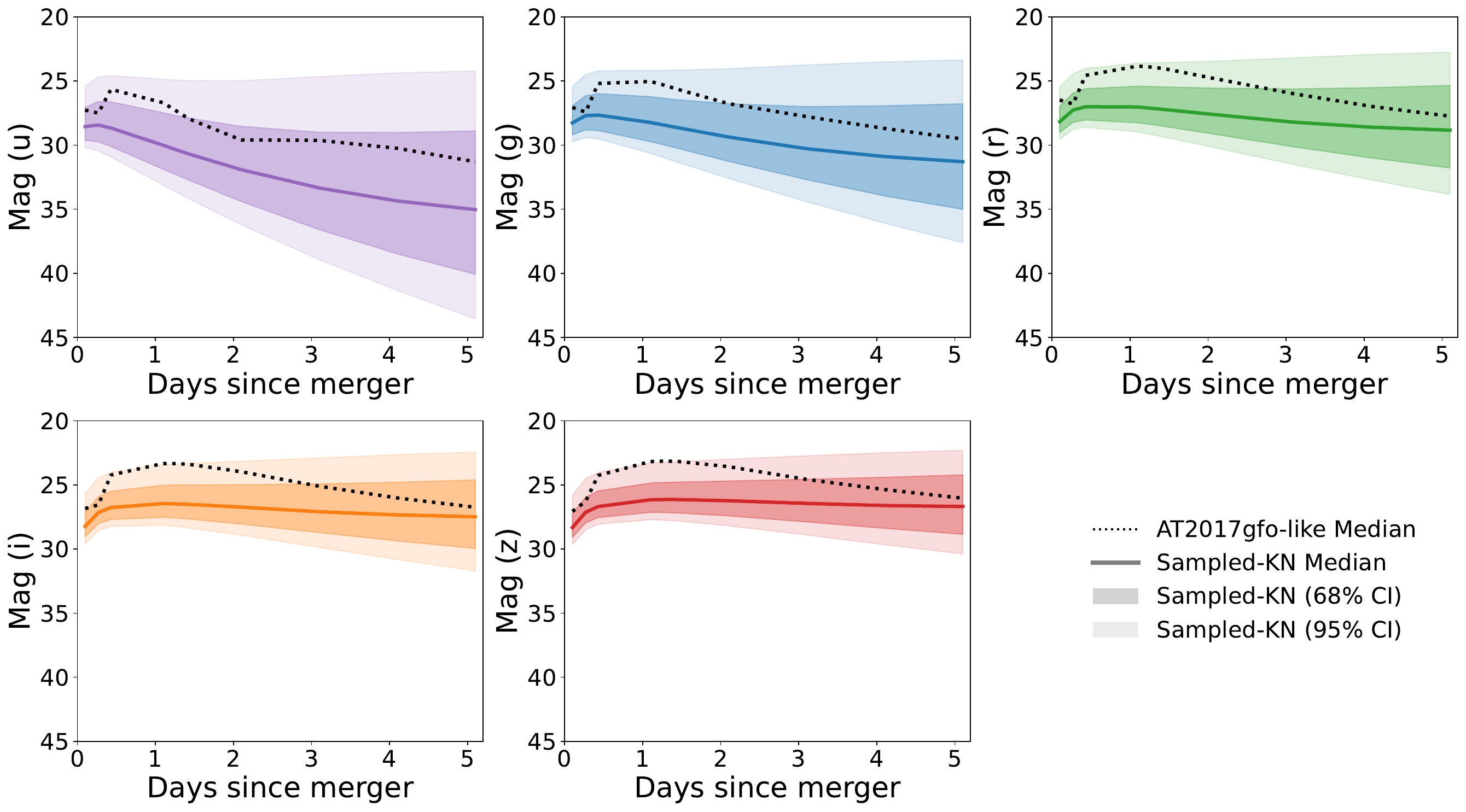}
\caption{Comparison of multi-band ($u,g,r,i,z$) light curves between AT2017gfo-like and sampled kilonova populations. The solid colored lines represent the median light curves of the sampled kilonova ensemble, with shaded regions indicating the $68\%$ and $95\%$ confidence intervals (CI). The black dotted line denotes the median light curves of the AT2017gfo-like samples. \label{fig:magnetarknmag}}
\end{figure*}

To evaluate the identification performance of magnetar kilonovae with the WFST, we conduct Monte Carlo simulations to generate $10,000$ samples, utilizing the afterglow parameters detailed in Table~\ref{tab:ag parameters} and magnetar kilonovae parameters detailed in Table~\ref{tab:kn parameters}. As illustrated in Figure~\ref{fig:magnetarknmag}, the median magnitude of AT2017gfo-like samples consistently resides above the median magnitude of the sampled magnetar kilonovae across all bands. We employed the same identification protocol described in Section~\ref{sec:identification framework}. However, because the magnetar kilonova model assumes a relatively simplified, isotropic ejecta structure with a uniform opacity, it effectively suppresses the spectral variations typically induced by distance-dependent redshift effects. This lack of distinct spectral information leads to a strong degeneracy between the distance and the amplitude A, which significantly increases the uncertainties in the estimation of A. Therefore, to ensure robust parameter inference, we impose a Gaussian prior with a 5\% uncertainty on the distance, which enables us to derive more accurate and reliable constraints. 

\begin{figure}[h!]
  \begin{subfigure}[h]{1\textwidth}
    \centering
    \includegraphics[width=\textwidth]{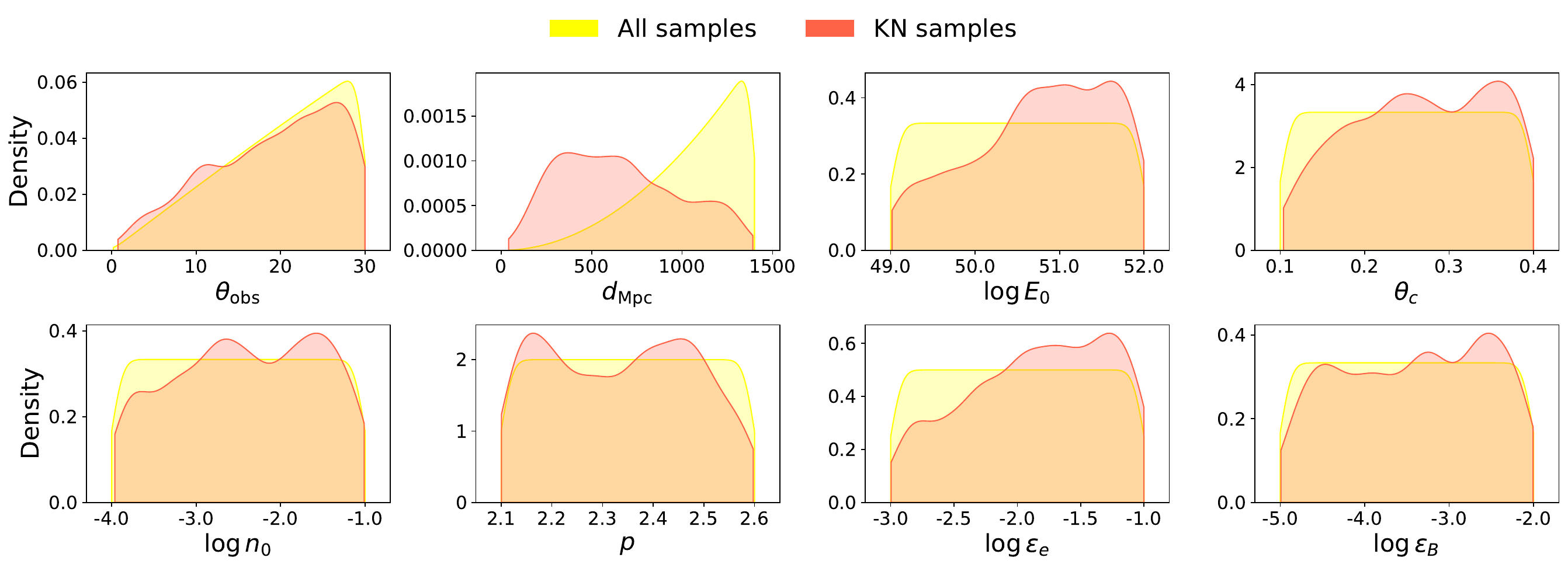}
    \caption{Parameter distributions \label{fig:a of magnetar}}
  \end{subfigure}  
  \par\medskip
  \begin{subfigure}[h]{1\textwidth}
    \centering
    \includegraphics[width=\textwidth]{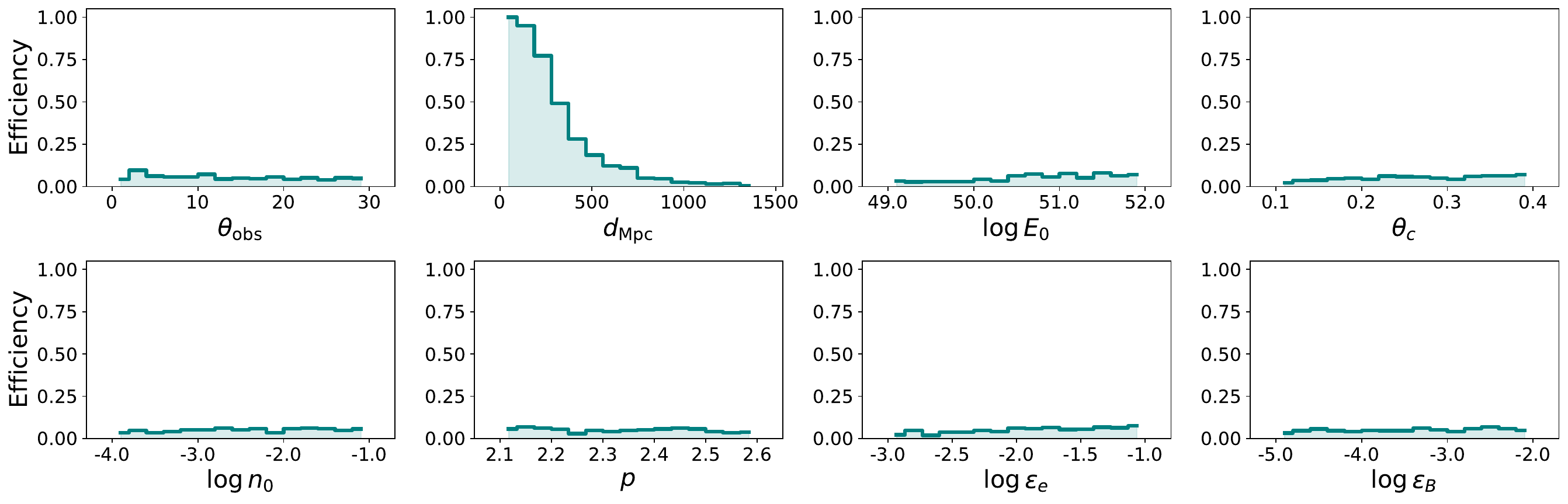}
    \caption{Identification efficiency \label{fig:b of magnetar}}
  \end{subfigure}
  \caption{Statistical analysis of afterglow parameters for magnetar kilonova samples. \textbf{(a):} Kernel Density Estimation distributions of afterglow parameters for the total simulated samples (yellow) and the kilonova-identified samples (red). \textbf{(b):} Identification efficiency across the parameter space. Parameters include: viewing angle $\theta_{\mathrm{obs}}$, distance $d_{\mathrm{Mpc}}$, jet energy $\log E_0$, half jet angle $\theta_c$, circumburst density $\log n_0$, electron index $p$, and energy fractions $\log \epsilon_e$ and $\log \epsilon_B$.} 
\end{figure}

Using this approach, we identify 506 magnetar kilonova samples. The parameter distributions for the afterglow and magnetar kilonova samples, shown in Figures~\ref{fig:a of magnetar}, are similar to our previous results of Case 2, which suggest a requirement for more luminous afterglows to achieve detection. Regarding the identification efficiency shown in Figure~\ref{fig:b of magnetar}, distance remains the dominant factor, while the efficiency curves for other physical parameters remain notably flat, further reinforcing our earlier conclusions. The primary distinction observed is that the identification efficiency exhibits a continuous decline as distance increases, a trend likely attributable to the intrinsically lower luminosity of the magnetar kilonova population compared to the samples analyzed previously. 

\end{document}